\documentclass[aps,prl,superscriptaddress,prbib,amsfonts,11pt]{revtex4}
\usepackage{amsmath,amssymb,graphicx}
\usepackage[utf8]{inputenc}
\usepackage{color}
\usepackage{lmodern}
\usepackage{float}
\usepackage[normalem]{ulem}
\usepackage{cancel}
\usepackage{epstopdf}
\newcommand{\CP}{{\mathbb{C}}{{P}}} 
\begin{document}
\title{Quartic metal: Spontaneous breaking of time-reversal symmetry due to four-fermion correlations in Ba$_{1-x}$K$_x$Fe$_2$As$_2$}

\author{Vadim Grinenko}
\email{vadim.a.grinenko@gmail.com}
\affiliation{Institute for Solid State and Materials Physics, Technische Universit\"at Dresden, 01069 Dresden, Germany}
\affiliation{Leibniz IFW Dresden, 01067 Dresden, Germany}
\author{Daniel Weston}
\affiliation{Department of Physics, KTH Royal Institute of Technology, SE-106 91 
Stockholm, Sweden}
\author{Federico Caglieris} 
\affiliation{Leibniz IFW Dresden, 01067 Dresden, Germany}
\author{Christoph Wuttke} 
\affiliation{Leibniz IFW Dresden, 01067 Dresden, Germany}
\author{Christian Hess} 
\affiliation{Leibniz IFW Dresden, 01067 Dresden, Germany}
\affiliation{Fakult\"at f\"ur Mathematik und Naturwissenschaften, Bergische Universit\"at Wuppertal, 42097 Wuppertal, Germany}
\author{Tino Gottschall}
\affiliation{Dresden High Magnetic Field Laboratory (HLD-EMFL) and W\"urzburg-Dresden Cluster of Excellence ct.qmat Helmholtz-Zentrum Dresden-Rossendorf, 01328 Dresden, Germany}
\author{Ilaria Maccari}
\affiliation{Department of Physics, KTH Royal Institute of Technology, SE-106 91 
Stockholm, Sweden}
\author{Denis Gorbunov}
\affiliation{Dresden High Magnetic Field Laboratory (HLD-EMFL) and W\"urzburg-Dresden Cluster of Excellence ct.qmat Helmholtz-Zentrum Dresden-Rossendorf, 01328 Dresden, Germany}
\author{Sergei Zherlitsyn}
\affiliation{Dresden High Magnetic Field Laboratory (HLD-EMFL) and W\"urzburg-Dresden Cluster of Excellence ct.qmat Helmholtz-Zentrum Dresden-Rossendorf, 01328 Dresden, Germany}
\author{Jochen Wosnitza}
\affiliation{Institute for Solid State and Materials Physics, Technische Universit\"at Dresden, 01069 Dresden, Germany}
\affiliation{Dresden High Magnetic Field Laboratory (HLD-EMFL) and W\"urzburg-Dresden Cluster of Excellence ct.qmat Helmholtz-Zentrum Dresden-Rossendorf, 01328 Dresden, Germany}
\author{Andreas Rydh}
\affiliation{Department of Physics, Stockholm University, SE-106 91 Stockholm, Sweden}
\author{Kunihiro Kihou} 
\affiliation{National Institute of Advanced Industrial Science and Technology (AIST), Tsukuba, Ibaraki 305-8568, Japan}
\author{Chul-Ho Lee} 
\affiliation{National Institute of Advanced Industrial Science and Technology (AIST), Tsukuba, Ibaraki 305-8568, Japan}
\author{Rajib Sarkar}
\affiliation{Institute for Solid State and Materials Physics, Technische Universit\"at Dresden, 01069 Dresden, Germany}
\author{Shanu Dengre}
\affiliation{Institute for Solid State and Materials Physics, Technische Universit\"at Dresden, 01069 Dresden, Germany}
 \author{Julien Garaud}
 \affiliation{Institut Denis Poisson CNRS-UMR 7013,  
 			 Universit\'e de Tours, 37200 France}
\author{Aliaksei Charnukha}
\affiliation{Leibniz IFW Dresden, 01067 Dresden, Germany}
\author{Ruben H\"uhne}
\affiliation{Leibniz IFW Dresden, 01067 Dresden, Germany}
\author{Kornelius Nielsch}
\affiliation{Leibniz IFW Dresden, 01067 Dresden, Germany}
\author{Bernd B\"uchner}
\affiliation{Institute for Solid State and Materials Physics, Technische Universit\"at Dresden, 01069 Dresden, Germany}
\affiliation{Leibniz IFW Dresden, 01067 Dresden, Germany}
\author{Hans-Henning Klauss}
\affiliation{Institute for Solid State and Materials Physics, Technische Universit\"at Dresden, 01069 Dresden, Germany}
\author{Egor Babaev}
\email{babaev.egor@gmail.com}
\affiliation{Department of Physics, KTH Royal Institute of Technology, SE-106 91 
Stockholm, Sweden}

\maketitle

Discoveries of ordered quantum states of matter are of great fundamental interest, and often lead to unique applications. The most well known example---superconductivity---is caused by the formation and condensation of pairs of electrons. A key property of superconductors is diamagnetism: magnetic fields are screened by dissipationless currents. Fundamentally, what distinguishes superconducting states from normal states is a spontaneously broken symmetry corresponding to long-range coherence of fermion pairs. 
Here we report a set of experimental observations in hole doped Ba$_{1-x}$K$_x$Fe$_2$As$_2$ which 
 are not consistent with conventional
 superconducting behavior. 
Our specific-heat measurements indicate the formation of fermionic bound states when the temperature is lowered from the normal state. However, for $x \sim 0.8$, instead of the standard for superconductors, zero resistance and diamagnetic screening, for a range of temperatures, we observe the opposite effect: the generation of self-induced magnetic fields measured by spontaneous Nernst effect and muon spin rotation experiments. The finite resistance and the lack of any detectable diamagnetic screening in this state exclude the spontaneously broken symmetry associated with superconducting two-fermion correlations. Instead, combined evidence from transport and thermodynamic measurements indicates that the formation of fermionic bound states leads to spontaneous breaking of time-reversal symmetry above the superconducting transition temperature.  These results  demonstrate the existence of a  broken-time-reversal-symmetry bosonic metal state. In the framework of  a multiband theory, such  a state is characterized by  quartic correlations: the long-range order exists only for {\it pairs} of fermion pairs.

\maketitle
\section{Introduction}

{The Bardeen-Cooper-Schrieffer (BCS) \cite{Bardeen1957a, Bardeen1957b} and Ginzburg-Landau \cite{Ginzburg1950} theories describe a superconducting state of matter arising via a single phase transition, from a symmetric normal state to a superconducting state that breaks $U(1)$ symmetry as a consequence of the formation and condensation of Cooper pairs of electrons.  Within the mean-field BCS theory, the condensate formation of  four-electron bound states is not  competitive with fermionic pair condensates. The situation is different in
superconductors   that break multiple symmetries.
Multicomponent states can, for example, be a  consequence of the presence of multiple electronic bands, where different components represent pairing in different bands. When the interactions between bands are frustrated, a  twofold degeneracy of the ground state can appear; such states are called $a_1\pm\mathrm{i}a_2$, where $a_\mathrm{j} = s$, $d$, $p$,... .} Mathematically this degeneracy is denoted as $Z_2$, and the total symmetry spontaneously broken by such a superconducting state is denoted $U(1)\times Z_2$. Such symmetry breaking occurs when phase differences between gaps in different bands have values other than $0$ or $\pi$ \cite{Stanev2010, Carlstrom2011b, Maiti2013, Boeker2017, Romer2019, Kivelson2020}. Such a state spontaneously breaks time-reversal symmetry (BTRS) since complex conjugation of the gaps (time reversal) leads to a different ground state. 

\begin{figure}
	\includegraphics[width=12cm]{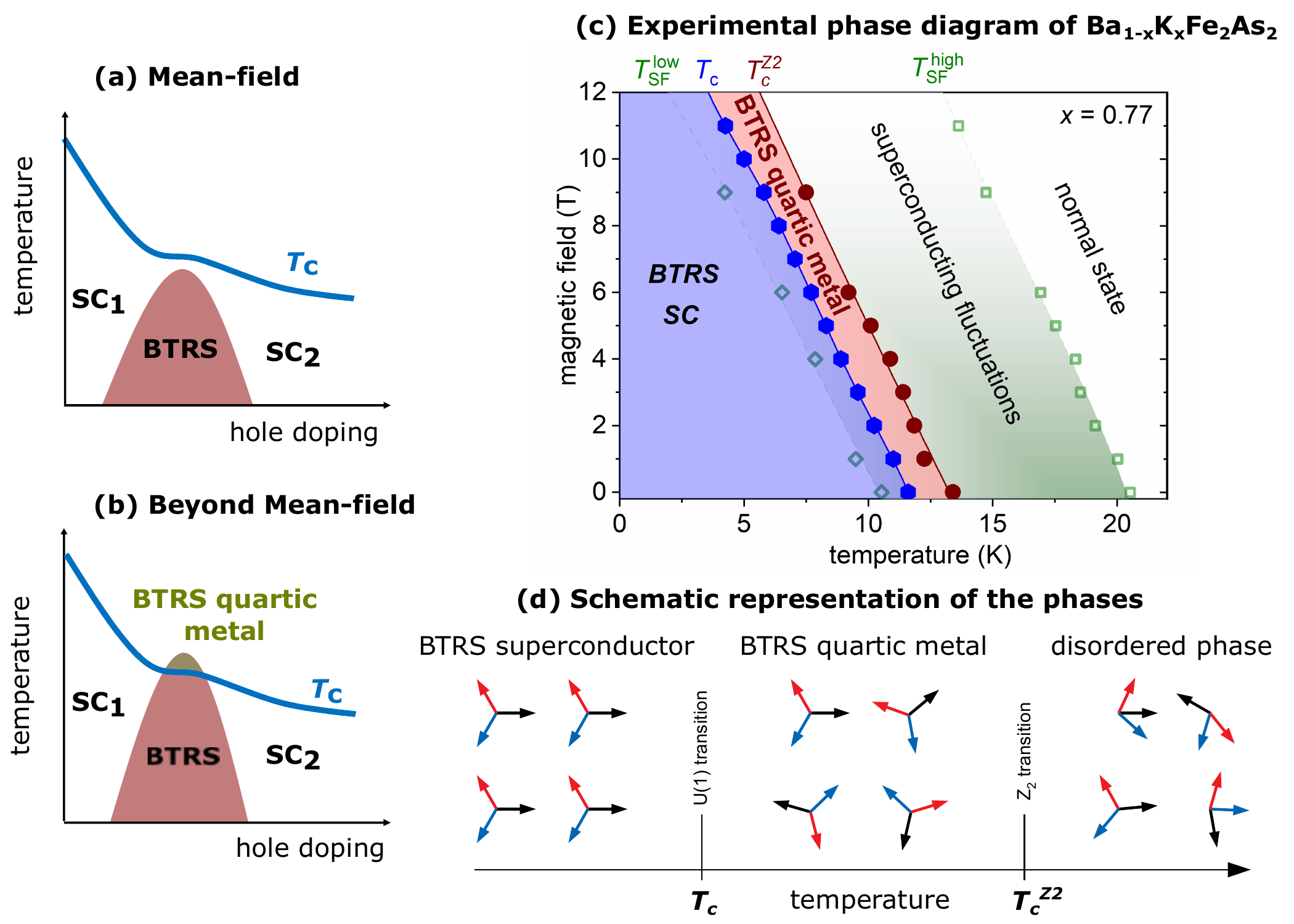}
	\caption{{\bf Phase diagrams. (a)} A schematic plot of the superconducting dome with two different $s_\pm$ states is separated by an intermediate state with a complex $s+\mathrm{i}s$ order parameter that breaks time-reversal symmetry. 
	{\bf (b)} The corresponding schematic phase diagram expected in the case when superconducting fluctuations are taken into account \cite{Bojesen2013, Bojesen2014a}. 
	{\bf (c)} Experimental magnetic-field phase diagram, showing (from left to right) a BTRS superconducting phase, a BTRS quartic metal phase that breaks time-reversal symmetry, the disordered phase with preformed Cooper pairs, and a normal metallic state. There are two characteristic crossover temperatures $T_{\rm SF}^{\rm low}$ and $T_{\rm SF}^{\rm high}$ showing the temperature range with superconducting fluctuations obtained from Nernst effect data and the temperature dependence of the resistivity. $T_{\rm c}$ is the superconducting transition temperature and corresponds to a zero-resistance state. $T_{\rm c}^{\rm Z2}$ is the $Z_2$ transition temperature at which time-reversal symmetry is spontaneously broken; it is obtained from the appearance of a spontaneous Nernst effect as shown in Fig.~\ref{Fig3}. {\bf (d)} Schematic illustration of the phase-locking configurations in different bands, denoted by arrows, corresponding to the   phases shown in panel {\bf (c)}.
	}
	\label{Fig1}
\end{figure}

In superconductors with broken time-reversal symmetry, at the level of mean-field theory, the $\mathrm{U}(1)$ symmetry is broken at a higher temperature $T_\mathrm{c}$ than the temperature $T_{\rm c}^{\rm Z2}$ at which the time-reversal $Z_2$ symmetry is broken. Within these models, states where $T_\mathrm{c} < T_{\rm c}^{\rm Z2}$ are forbidden and for $s+\mathrm{i}s$ and $s+\mathrm{i}d$ states some fine tuning or special symmetry is in general required to obtain degenerate temperatures $T_\mathrm{c} = T_{\rm c}^{\rm Z2}$ \cite{Stanev2010, Carlstrom2011b, Maiti2013, silaev2017phase, Boeker2017, Grinenko2020}. Therefore, a phase diagram as a function of doping at mean-field level generically looks like a dome of the $a_1+\mathrm{i}a_2$ state between two superconducting states 
having $a_1$ and $a_2$ order parameters, where the maximal $T_{\rm c}^{\rm Z2}$ does not reach $T_\mathrm{c}$ (Fig.~\ref{Fig1}a).

{ The situation is different if one considers multicomponent systems beyond mean-field approximation, where new states of matter may form with fermion quadrupling, preceding superconducting state \cite{babaev2004superconductor}}. 
Fluctuation effects have been considered in the London model of 
$U(1)\times Z_2$
superconductors \cite{Bojesen2013, Bojesen2014a, Carlstrom2015}, and it was found that in both two and three spatial dimensions a situation can arise where $T_\mathrm{c} < T_{\rm c}^{\rm Z2}$.
This implies the existence of a novel bosonic metallic state, where time-reversal symmetry is spontaneously broken. 
 The novel bosonic metallic state is separated from the normal state by a second phase transition, at which time-reversal symmetry is restored. Therefore, beyond mean-field approximation, the line of $T_{\rm c}^{\rm Z2}$ as a function of doping can appear above the line of $T_\mathrm{c}$ (Fig.~\ref{Fig1}b) \cite{Bojesen2013, Bojesen2014a}.  This is a counterpart of  the metallic superfluid state, sought after in ultra-high pressure physics,
and dense nuclear matter \cite{babaev2004superconductor, Smorgrav2005b}, with the key difference that here the interband coupling only allows the breaking of a discrete $Z_2$ symmetry. Other recently discussed examples of a sequence of phase transitions with fluctuation-induced intermediate phases  
were proposed, for example, in the context of pair-density-wave superconductors \cite{agterberg2008dislocations,berg2009charge}, loop-current superconducting models \cite{brydon2019loop}, and nematic states \cite{cho2020z}.

{ A multicomponent superconducting $s+\mathrm{i}s$ state, in which time-reversal symmetry is broken, has been theoretically expected in Ba$_{1-x}$K$_x$Fe$_2$As$_2$ \cite{Stanev2010, Carlstrom2011b, Maiti2013, Boeker2017}.} 
Recent experimental evidence of such an $s+\mathrm{i}s$ state emerging in this material at $x \sim 0.8$ were obtained in muon spin rotation ($\mu$SR) experiments \cite{Grinenko2017, Grinenko2018}, which were focused primarily on low-temperature states. The initial data indicated that the dome of the $s+\mathrm{i}s$ phase reaches the superconducting $T_{\rm c}$.
{As mentioned  above, if fluctuations were negligible,
that would require a physically highly unlikely fine-tuning \cite{Maiti2013}. 
Monte-Carlo calculations beyond mean-field approximation in the London model yield two possible scenarios. In the first scenario, $T_\mathrm{c}$ merges with $ T_{\rm c}^{\rm Z2}$ into a single transition that exists for a range of dopings \cite{Bojesen2014a}. However, the merged transition is first order, and no signs of first-order transitions have been detected in Ba$_{1-x}$K$_x$Fe$_2$As$_2$. This calls for the investigation of the second scenario, in which the critical temperature $T_{\rm c}^{\rm Z2}$ exceeds $T_\mathrm{c}$  \cite{Bojesen2013, Bojesen2014a}.}

We conducted combined thermodynamic as well as thermal and electrical transport measurements, and    find  agreement with scenario (b), namely that Ba$_{1-x}$K$_x$Fe$_2$As$_2$ realizes a novel type of metallic state of matter. This state breaks time-reversal symmetry above $T_\mathrm{c}$  and is characterized by an order parameter that is fourth order in fermionic fields. For brevity, we will refer to this state as ``BTRS quartic metal''. Our key experimental findings for the samples in the quartic metal phase are:
the observation of a spontaneous Nernst effect and enhanced muon spin relaxation rate, which  indicate a BTRS state  in the restive state above the superconducting transition temperature $T_{\rm c}$ accompanied
by a lack of a diamagnetic response at $T_{\rm c}^{\rm z2}$. The presence of the phase transition at $T_{\rm c}^{\rm z2}$ is consistent with the anomalies in the specific heat and ultrasound data above $T_{\rm c}$. No such anomalies exist in the reference samples. The beyond-mean-field nature of the quartic phase is supported by 
the observation of pairing fluctuations setting in below 2~$T_{\rm c}$, obtained from measurements of thermoelectric and electric transport properties. The resulting experimental phase diagram of the sample with the BTRS quartic phases is shown in Fig.\ref{Fig1}c  
We provide a theoretical analysis of the occurrence of the BTRS quartic metal phase and some of its properties in the methods section (Figs.~\ref{FigDW} and \ref{FigED3}) and the supplementary information.

\section{Results}

\begin{figure}
	\includegraphics[width=\columnwidth]{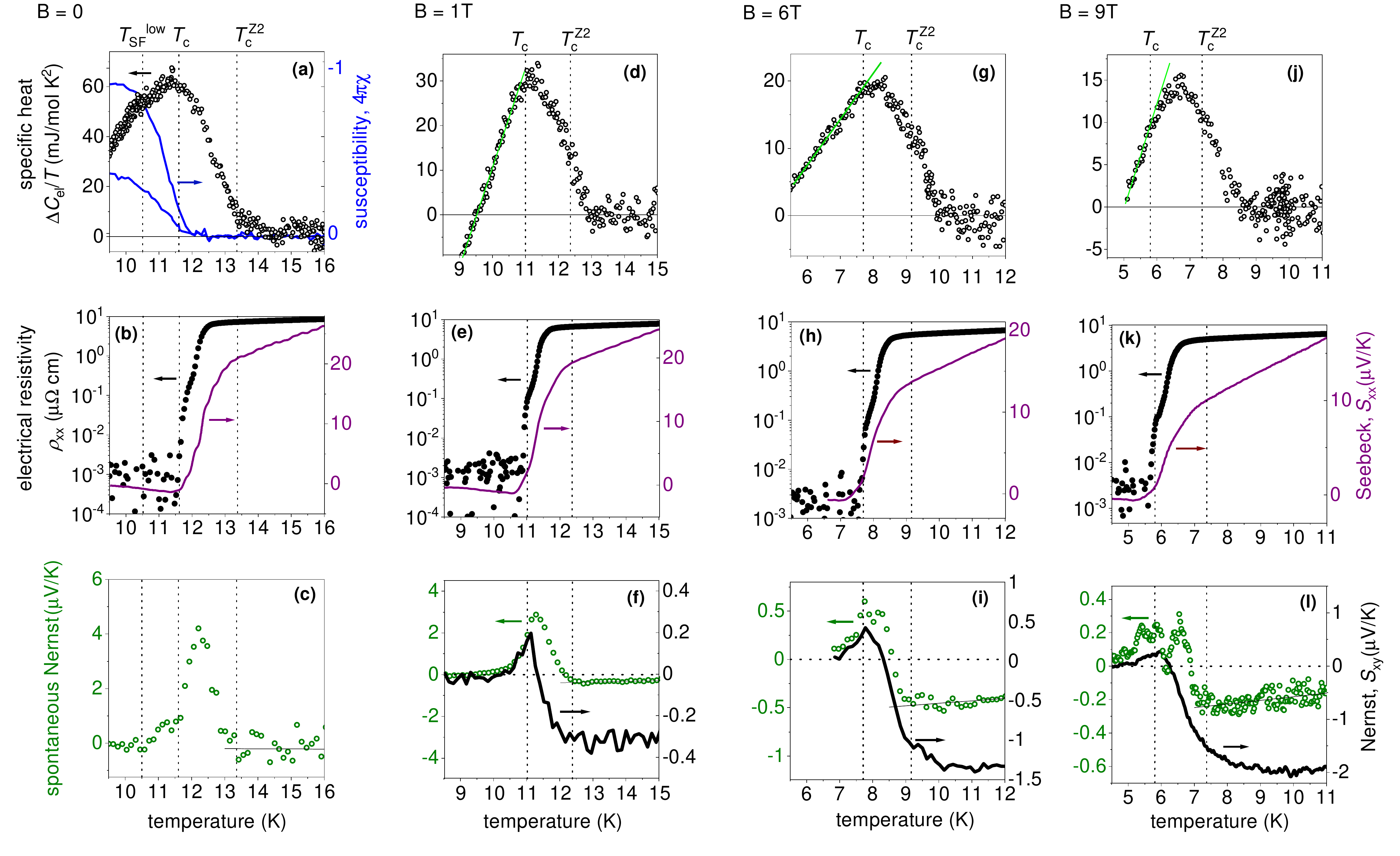}
	\caption{{\bf Characteristic temperatures.}  Temperature dependencies of various physical properties measured in different magnetic fields for the single crystal with $x = 0.77$ reveal the existence of several characteristic temperatures. {\bf (a)} The temperature dependence of the zero-field high-resolution specific heat $\Delta C_{\rm el}/T$ and the static magnetic susceptibility measured in $B\parallel ab = 0.5$~mT shows splitting between $T_{\rm c}$ and $T_{\rm c}^{\rm Z2}$ {\bf (b)} The appearance of the diamagnetic response in the susceptibility at $T_{\rm c}$ corresponds to zero resistance and a minimum in the Seebeck coefficient. {\bf (c)} The onset of the specific-heat anomaly at $T_{\rm c}^{\rm Z2}$ is accompanied by the appearance of a strong spontaneous Nernst effect as a consequence of a BTRS state. The spontaneous Nernst signal goes to zero at $T_{\rm SF}^{\rm low}$ deep in the superconducting state when superconducting fluctuations become negligible. {\bf (d-l)} In applied magnetic field the characteristic temperatures decrease. $T_{\rm c}^{\rm Z2}$ shifts further away from the onset of the specific-heat anomaly [panels (d, g, j)]. The superconducting $T_{\rm c}$ corresponding to zero resistance [panels (e, h, k)] reduces below the maximum in the specific heat [panels (g, j)].
	The resistance starts dropping significantly at $T_{\rm c}^{\rm Z2}$ due to the increased density of incoherent preformed pairs.} In turn, the odd Nernst effect shows a maximum at $T_{\rm c}$ indicating vortex-lattice melting [panels (f, i, l)].
	\label{Fig3}
\end{figure}

 The main experimental results indicating the occurrence of the BTRS quartic metal phase are summarized in Fig.~\ref{Fig3}. In this figure we compare the results of 
high-resolution measurements of the specific heat, (using AC calorimetry), 
electric and thermoelectric transport studies of Ba$_{\rm 1-x}$K$_{\rm x}$Fe$_2$As$_2$ single crystals with $x = 0.77$ (details of the techniques are described in the methods section). 
In zero magnetic field, we found a splitting between $T_{\rm c}$ defined by zero resistance and the onset temperature of the 
specific-heat anomaly [panels (a, and b)]. Importantly, we observed the appearance of a spontaneous Nernst signal right at the onset specific-heat anomaly [panel (c)]  obtained as described in the methods section Fig.\ref{FigED2_b}. 

Until now, a spontaneous Nernst effect has been observed only in a few superconductors. It can have several different origins. In La$_{2-x}$Ba$_x$CuO$_4$ it was observed close to $x = 1/8$  \cite{Li2011, Soumyanarayanan2016} together with non-zero polar Kerr effect \cite{Karapetyan2012}, which was interpreted as evidence for broken time-reversal symmetry. Alternatively, it has been shown that a polar Kerr effect and a spontaneous Nernst signal can appear due to  chiral charge ordering in the pseudogap phase with preserved-time reversal symmetry \cite{Hosur2013}. Recently, a spontaneous Nernst effect was observed in the vortex-liquid phase of Fe$_{1+y}$Te$_{1-x}$Se$_x$ \cite{Chen2020}, where it was attributed to a subtle interplay of $s$-wave superconductivity and interstitial magnetic Fe impurities. In contrast, our Ba$_{\rm 1-x}$K$_{\rm x}$Fe$_2$As$_2$ single crystals show a very small amount of magnetic impurities \cite {Grinenko2018}. Also, there is no evidence for charge ordering in Ba$_{\rm 1-x}$K$_{\rm x}$Fe$_2$As$_2$. This points to the spontaneous Nernst signal being related to the BTRS superconductivity found in $\mu$SR experiments for this doping level (Fig.~\ref{muSR_ED}). 

The situation is very different in single crystals not showing the BTRS state. For the sample with $x = 0.54$, there are only some weak noisy features in the signal shown in Fig.~\ref{Fig4}a, which cannot be clearly distinguished from noise in the raw data (Fig.~\ref{FigED2}a). However, we note that reentrant transitions giving such effects in $s+is$ superconductors have been predicted theoretically \cite{Carlstrom2015}, which calls for further investigation of  whether these tiny features are indeed pure noise. For $x = 1$, we detect no spontaneous Nernst signal. 

If an external field is applied, both $T_{\rm c}$ and $T_{\rm c}^{\rm Z2}$ move to lower temperatures, consistent with $T_{\rm c}^{\rm Z2}$ being related to Cooper pairing.  The broad specific-heat anomaly changes shape and a kink-like feature appears slightly below the onset temperature of the anomaly [panels (d, g, and j)].  The feature in the specific heat coincides with the onset of a strong spontaneous Nernst signal and the onset of the positive contribution in the {\it conventional} Nernst signal (odd in magnetic field) [panels (f, i, and l)]. In high magnetic fields, $T_{\rm c}$ is suppressed below the temperature of the maximum of the specific heat.  However, the expected second anomaly at $T_{\rm c}$ is barely visible [panels (g, and j)] due to disorder effect naturally present in the doped samples that broadens the vortex-lattice melting transition. We note that the only crucial requirement to characterize the state as $Z_2$ phase is lack of superconductivity, i.e. long-range order in Cooper pairs. It is not principally important whether one has a sharp vortex lattice melting transition. However,  the pronounced peak in the conventional Nernst signal [panels (f, i, and l)] is 
supportive for a relatively sharp  melting of the vortex lattice at $T_{\rm c}$.  
Irrespective of the nature of the vortex state, the data in a magnetic field is consistent with  
two distinct phase transitions: a $T_{\rm c}$, above which at least some vortices are mobile, and  $T_{\rm c}^{\rm Z2}$, where $Z_2$ symmetry associated with the interband phase difference  is broken. 
The details of the in-field behaviour of the dominant specific-heat anomaly are shown in Fig.~\ref{ED_0p8_PD}. In zero field, $T_{\rm c}^{\rm Z2}$ is located close to the onset of the dominant anomaly 
in the specific heat according to thermoelectric probes as seen from comparison of panels (a) and (c) in Fig.~\ref{Fig3}. The large size of the mean-field contribution in zero field and its possible rounding close to the onset make it difficult to 
detect the fluctuation-induced  
 $Z_2$ transition in the specific heat. However, we found an extra support for the anomaly at the ultrasound measurements shown in the supplementary information Fig.\ref{ultrasound_chi_ED}.

To verify that superconducting fluctuations are present above $T_{\rm c}$ we measured the conventional Nernst effect. We present the data in Fig.~\ref{Fig4} and in the methods section. Summarising the transport data for different samples, we conclude that in Ba$_{\rm 1-x}$K$_{\rm x}$Fe$_2$As$_2$ at $x \sim 0.8$ strong superconducting fluctuations occur in a broad temperature range, i.e.\ the phase ordering temperature is significantly suppressed relative to the pairing temperature. In contrast, the samples with other doping levels do not show such behavior. This is consistent with the fact that this doping  corresponds to the maximum of the BTRS dome (Fig.~\ref{Fig2}i), which in turn corresponds to maximally frustrated interband coupling. This leads to (i) suppression of ordering temperatures \cite{Bojesen2014a} and (ii) the occurrence of the BTRS quartic metal phase above $T_{\rm c}$.

To determine the doping range in which the BTRS quartic phase can occur, we 
investigated Ba$_{\rm 1-x}$K$_{\rm x}$Fe$_2$As$_2$ single crystals with $0.55 \lesssim x \le 1$, covering a region in the phase diagram with two different $s\pm$ and intermediate $s+\mathrm{i}s$ states \cite{Grinenko2018,Cho2016}.
The zero-field specific heat and the low-field static magnetic susceptibility close to $T_\mathrm{c}$ for Ba$_{\rm 1-x}$K$_{\rm x}$Fe$_2$As$_2$ single crystals with different doping levels are shown in Fig.~\ref{Fig2}. The onset of a strong diamagnetic signal in the susceptibility indicates the appearance of screening currents. Close to
this temperature ($T_\mathrm{c}$) the electrical resistivity goes to zero as shown in Figs.~\ref{Fig3}, ~\ref{Fig2}, and \ref{ED_SH_S14} indicating superconductivity with a long-range phase-coherent state.

Away from the doping value $x \sim 0.8$, we observed consistent $T_\mathrm{c}$ in the susceptibility and specific-heat (Fig.~\ref{Fig2}). However, the picture becomes strikingly different for the samples with a doping level $x \sim 0.8$ (Figs.~\ref{Fig3}a, ~\ref{Fig2}d, and~\ref{ED_SH_S14}a). 
Similar to the sample with $x = 0.77$, for all samples with $x \sim 0.8$
the specific-heat jump in zero field occurs at a temperature 
above 
any detectable diamagnetic response (magnification of the ac magnetization data by $10^3$ is shown in Fig.~\ref{Fig2}f). 
In contrast, the commonly observed deviations between resistive and calorimetric probes are in the opposite direction: for inhomogeneous samples, some areas start contributing to the diamagnetic response at temperatures higher than that of a discernible specific-heat jump (see also the methods section). For the samples with $x \sim 0.8$, the onset of the strong diamagnetic response at $T_{\rm c}$ occurs close to the maximum in the specific-heat jump, indicating that Cooper-pair formation has already occurred in most of the volume.

\section{Discussion}

In the method section, we discuss the mechanism resulting in the spontaneous Nernst effect  and thermodynamic signatures in the BTRS quartic metal phase. Our set of measurements demonstrates
the existence of the observed in Monte-Carlo calculations new fermion-quadrupling state above $T_{\rm c}$. Namely  a $Z_2$-condensate with long-range four-fermion correlations.

The experimental results for the sample with the BTRS quartic metal phase are summarized in the phase diagram in Fig.~\ref{Fig1}c. At low temperatures there is a superconducting phase. It has an $s+{\rm i}s$ order parameter that breaks $U(1)\times Z_2$ symmetry, as demonstrated in $\mu$SR experiments \cite{Grinenko2018}. A simple three-band model describing this state is illustrated in the bottom panels of Fig.~\ref{Fig1}d. The arrows represent the phases of the superconducting gaps in different bands. In the superconducting state (left panel) the phases of the superconducting gaps in different bands are ordered, and there are two energetically equivalent phase configurations corresponding to domains with $s+{\rm i}s$ and $s-{\rm i}s$ states. Increasing the temperature above the characteristic temperature $T_{\rm SF}^{\rm low}$, 
we detect the appearance of superconducting phase fluctuations
resulting in a gradual increase of the conventional Nernst coefficient (Fig.~\ref{Fig3}, bottom row). The field dependence of $T_{\rm SF}^{\rm low}$ shown in Fig.~\ref{Fig1} is obtained from the Nernst effect. At $T_{\rm c}$ superconductivity disappears and $U(1)$ symmetry is restored. In finite fields the resistive transition is associated with vortex-lattice melting or onset of  mobility of some of vortices in a system with pinning. 
When the system enters a resistive state, the conventional Nernst effect in applied magnetic field shows a sharp peak the standard signature of vortices motion \cite{Behnia2016}. The line corresponding to the superconducting critical temperature $T_{\rm c}$ in Fig.~\ref{Fig1} is plotted using the field dependence of the temperature at which the resistance disappears.  In turn, $T_{\rm c}^{\rm Z2}$ is determined from the spontaneous Nernst signal.  

Additional signature of  the unconventional character of the state 
above the superconducting phase transition comes from the ultrasound measurements shown in the supplementary information Figs.~\ref{ultrasound_chi_ED}, and ~\ref{ultrsound_sup_0p81}.
The  ultrasound, as a complimentary probe  of  thermodynamic properties, is expected  be sensitive to the $Z_2$ phase transition  especially because of the the non-trivial hybridization of phase-difference and amplitude modes in  $s+is$ superconductors
\cite{Carlstrom2011b,Maiti2013,garaud2018properties}.
The experimental ultrasound data shows an additional 
anomaly at the temperature, which was interpreted as $Z_2$ phase transition in the electrical and thermal transport experiments for the samples with $x \sim 0.8$. By contrast, the ultrasound measurements in the reference samples  show only an anomaly located at superconducting $T_{\rm c}$. 

Extensive theoretical analysis of our experimental data (see the methods section and the supplementary information) allows us to conclude that above the resistive phase transition at $T_{\rm c}$ there are fluctuating pairs, but these fluctuations are correlated in different bands: the interband phase differences remain locked by interband Josephson coupling  up to $T_{\rm c}^{\rm Z2}$, as illustrated in the bottom central panel of Fig.~\ref{Fig1}d. Because the interband Josephson coupling is frustrated, there are two energetically equivalent phase-locking patterns, and thus this state spontaneously breaks $Z_2$ symmetry but preserves $U(1)$ symmetry, resulting in the formation of a novel state. The spontaneously broken symmetry is related to interband phase differences. In other words, the corresponding order parameter is proportional to the real part of the product of the gap function in one band and the complex conjugate of the gap function in another band. Therefore, it depends only on the interband phase differences but not on the absolute value of the superconducting phase. The new type of metallic phase above $T_{\rm c}$ is characterized by time-reversal-symmetry-breaking correlations that are quartic in fermionic fields, without nontrivial correlations in quadratic terms. In the quartic phase we find a strong spontaneous Nernst signal
accompanied by the enhancement of the muon spin relaxation rate above $T_{\rm c}$ (Fig.~\ref{muSR_ED}).
This is consistent with the theoretical model because spontaneous magnetic fields in the $s+{\rm i}s$ state originate from interband phase-difference and relative density gradients \cite{garaud2014domain,Grinenko2018}, and thus they are expected to become somewhat stronger in the quartic metal phase due to the absence of Meissner screening. 

At $T_{\rm c}^{\rm Z2}$, domain walls in the phase differences proliferate and time-reversal symmetry is restored, eliminating spontaneous magnetic fields and unconventional thermoeletric effects. However, the evidence for the presence of incoherent pairing fluctuations  remains up to a  much higher temperature. This results in a deviation from the normal-state behavior of the resistivity and the conventional Nernst effect (Fig.~\ref{Fig4}). Above the crossover temperature $T_{\rm SF}^{\rm high}$ the system is in the normal state (Fig.~\ref{Fig1}) without detectable pairing fluctuations.

While the original Cooper pairing mechanism \cite{Cooper1956} is not directly generalizable to four-fermionic condensates, the states are possible in multicomponent systems. The rapidly growing family of superconductors with multiple broken symmetries 
and with significant fluctuations effects \cite{Yamashita2015},  suggests that this kind of a state may not be rare.  In particular, the phase diagrams of PrPt$_4$Ge$_{12}$ and PrOs$_4$Sb$_{12}$ families of filled skutterudite alloys share similarities with the Ba$_{\rm 1-x}$K$_{\rm x}$Fe$_2$As$_2$ system \cite{Shu2011, Zhang2015, Zhang2019a}. 
 Another promising material is a recently discovered $p$-wave candidate UTe$_2$ \cite{Ran2019, Metz2019, Hayes2020}.
We note that
the $Z_2$ phases are much more common and expected to be observed in a broader temperature range 
in two-dimensions \cite{Bojesen2013}. 
That suggests   
 carrying out similar studies in thin films made of  BTRS superconductors or 2D materials with BTRS states, such us one predicted in twisted bilayer graphene \cite{Chichinadze2020, Gonzalez2020}.
By the same token, quasi-one dimensional samples of BTRS superconductors should form quite generically  $Z_2$   quartic phases, since one cannot break a continuous symmetry in one dimension.

The unique properties that we observe open up questions regarding further fundamental properties and of the state we observe and its possible application. As discussed in the Methods section, its effective model is a Skyrme-type model  (see the methods section)  that indicates that the state has nontrivial topological excitations, which calls for probing these states in scanning SQUID experiments. The strong magnetic response to thermal gradient, unaffected by Meissner screening, raises the question of the potential for utilizing this state in sensors.

The state  we discussed is   one in a growing family of theoretically proposed 
  fluctuation-induced four-fermionic orders (see e.g. \cite{babaev2004superconductor,agterberg2008dislocations,berg2009charge,brydon2019loop}.) This suggests carrying a comparative experimental study with similar types of measurements in other candidate materials. For example, thermoelectric and ultrasound probes may be carried out to identify the four-fermion order anticipated to form at ultrahigh compression in hydrogen, deuterium and hydrides \cite{babaev2004superconductor,babaev2005observability}.

\section{Appendix}
\renewcommand{\theequation}{M\arabic{equation}}
\renewcommand{\thefigure}{A\arabic{figure}}
\renewcommand{\thetable}{A\arabic{table}}
\setcounter{equation}{0}
\setcounter{figure}{0}
\setcounter{table}{0}

\subsection{Theoretical analysis}

In the  single-component weak-coupling mean-field BCS theory, the phase transition separates a fermionic normal state from a bosonic superconducting state described by a classical field that is proportional to the complex gap function $\Delta$. Going beyond mean-field theory, one finds that at weak coupling there are Cooper-pairing fluctuations in the normal state above the critical temperature \cite{aslamazov1968effect}, while at stronger coupling no fermionic pair-breaking occurs at the phase transition \cite{leggett1980diatomic, nozieres1985bose, emery1995importance}. Unless a superconductor is strongly type I, the phase transition is driven by topologically nontrivial phase fluctuations \cite{Peskin1978, Dasgupta1981}. In zero external field, these fluctuations are described as proliferation of vortex loops, whereas in finite fields they are characterized as vortex-lattice melting \cite{Peskin1978, Dasgupta1981, nelson1988vortex, fisher1991thermal, Svistunov2015}. Although this implies that the normal state just above the phase transition has short-ranged bosonic fluctuations, this state does not constitute a separate phase since it lacks long-range order and is thus not distinguished by symmetry from a simple metallic state. Rather, with increased temperature the short-ranged bosonic correlations gradually vanish without a phase transition.

In multi-component  superconductors that break time-reversal symmetry there are multiple transitions. At the level of mean-field theory, the superconducting phase transition always occurs at a temperature equal to or higher than the BTRS transition temperature. All previously known experimental results fit the mean-field picture, in which the superconducting phase transition always occurs at a temperature equal to or higher than the BTRS transition temperature. 
In Ba$_{\rm 1-x}$K$_{\rm x}$Fe$_2$As$_2$, we observe the opposite behavior: 
$T_\mathrm{c}^{\rm Z2}$  exceeding $T_\mathrm{c}$. 
We first theoretically investigate the scenario where  the system can be described by a Ginzburg-Landau model with fluctuation corrections, which we include via Monte-Carlo simulation. Physically, this assumes a situation where the superconducting critical temperature is close to the weak-coupling mean-field critical temperature and fluctuations play a role only for a narrow temperature range. Under this assumption, one can expand in powers of the order parameter and its gradients, retain only the lowest-order terms and include fluctuations in the resulting GL model. Ginzburg-Landau models for such systems have been derived microscopically, see e.g.\ Refs.~\cite{Maiti2013, Garaud2017}. However, quantitative certainty about the form of the Ginzburg-Landau model requires much deeper insight into the microscopic physics of the material, which is currently lacking. We therefore begin by including fluctuations in the simplest BTRS form of the Ginzburg-Landau model consistent with Refs.~\cite{Maiti2013, Garaud2017}, and use the experimental observations to put constraints on this model (further details can be found in the method section and the supplementary information).

We first consider the Ginzburg-Landau model for a clean three-band superconductor in three spatial dimensions given by the free-energy density
\begin{equation}
  f = \tfrac{1}{2}(\nabla \times \mathbf{A})^2
      + \sum_i \tfrac{1}{2} |(\nabla + \mathrm{i}e\mathbf{A}) \psi_i|^2
      + a_i |\psi_i|^2 + \frac{b_i}{2} |\psi_i|^4
      + \sum_{i<j} \eta_{ij}|\psi_i| |\psi_j| \cos(\phi_i-\phi_j),
  \label{contf}
\end{equation}
which has been argued to describe the material in question \cite{Maiti2013, Garaud2017}. 
Here, $\mathbf{A}$ is the magnetic vector potential and $\psi_i=|\psi_i|^{i\phi_i}$ are matter fields corresponding to the superconducting components. There is also a reduced version of this model with only two components but a biquadratic Josephson coupling $\psi_1^2\psi_2^*{}^2+c.c.$ \cite{garaud2016thermoelectric}; our conclusions will apply to this case as well. Note however that the derivation of the Ginzburg-Landau functional (\ref{contf}) neglects some terms, such as mixed gradient  terms coming from Fermi-liquid corrections  or strong correlations
~\cite{leggett1975theoretical, Sjoberg:76, kuklov2004superfluid,Kuklov2006, Svistunov2015, Sellin.Babaev:18}. These terms play an important role  for the existence and size of the quartic metal state, since in general they affect the energy cost of domain-wall excitations relative to that of vortex excitations.

Apart from terms present in an ordinary single-component Ginzburg-Landau model, we have Josephson-coupling terms that directly couple the three components. Depending on the sign of the coefficient $\eta_{ij}$, a Josephson term will be minimized when the corresponding phase difference is either $0$ or $\pi$. For some combinations of signs of the coefficients $\eta_{ij}$, it is impossible to simultaneously minimize each term, and in these cases the system is frustrated. If this phase frustration is strong enough, the ground state will be such that the phase differences between the components are not all $0$ or $\pi$. This defines $s+\mathrm{i}s$ superconductivity where time-reversal symmetry is broken \cite{Stanev2010, Carlstrom2011b, Maiti2013}. There are four combinations of signs of the coefficients $\eta_{ij}$: two equivalent combinations that give rise to phase frustration and two equivalent combinations that do not. (The equivalence consists of switching the sign of the couplings related to a certain phase and adding $\pi$ to this phase.)
 
For definiteness and without loss of generality we consider the case when each Josephson coupling is repulsive, meaning that all phase differences tend to be equal to $\pi$. Since the $Z_2$ symmetry is broken when the phase differences are not all equal to $0$ or $\pi$, we can define a $Z_2$ Ising order parameter $m$.
We define $m$ to be equal to $+1$ for one of the chiralities of the phases and equal to $-1$ for the other chirality (Fig.~\ref{FigED3}). In other words, the field $m(\mathbf{r})$ takes the value $+1$ at points where the phases are ordered like 1, 2, 3, and takes the value $-1$ at points where the phases are ordered like 1, 3, 2. The chirality is thus determined by the phase differences of the gaps in different bands, $\phi_i-\phi_j = \arccos{\frac{1}{2}[\Delta_i\Delta_j^* + \Delta_i^*\Delta_j]}$, and therefore the state is characterized by an order parameter that is fourth order in fermionic fields. When the $Z_2$ symmetry is broken, the spatial average $m$ of the field $m(\mathbf{r})$ is nonzero.
Our main goal is to consider the effects of nonzero external magnetic field and density fluctuations. 

Our theoretical analysis, which goes beyond mean-field approximation, provides an interpretation of the experimental results. In this picture, two phase transitions that are distinct by symmetry can exist in a multiband BTRS superconductor, even in the presence of density fluctuations. There is a separate $Z_2$ transition that appears above the superconducting transition, and thus there is a state in which the superconducting phase is disordered but the phase differences between bands have nontrivial values different from $0$ or $\pi$. This is a novel metallic state that spontaneously breaks time-reversal symmetry. It is distinguished by an Ising-type $Z_2$ order parameter, which describes order associated with interband phase differences. In contrast to the conventional mechanisms for time-reversal symmetry breaking due to magnetism, here the time-reversal symmetry breaking originates purely from momentum space, i.e.\ from interband Josephson-like currents (note that an external magnetic field does not break that symmetry explicitly). Therefore, the novel state is a resistive metallic state that has preformed Copper pairs and persistent interband Josephson-like currents. The two $Z_2$ states correspond to two different Josephson-current ``loops" in momentum space: either from the first component to the second, from the second to the third and from the third back to the first, or in the opposite order (for an illustration see Fig.~\ref{FigED3}). For a uniform state this is a purely momentum-space phenomenon, i.e.\ it may be viewed as chirality in momentum space or order-parameter space. It does not involve real-space currents and thus does not involve spontaneous magnetic fields. However, in the presence of defects and boundaries there appear  interband phase-difference gradients, resulting in real-space currents and spontaneous magnetic fields \cite{garaud2014domain}.
This, as discussed below, persists in the non-superconducting  time-reversal-symmetry-breaking quartic metal state, giving rise to the observed thermoelectric phenomena.
The novel state represents a quartic bosonic metal with broken $Z_2$ symmetry.

\subsubsection{Effective Skyrme model and spontaneous magnetic fields in quartic metal}

As shown in Ref.~\cite{garaud2014domain}, there are spontaneous magnetic fields in the $s+\mathrm{i}s$ state. These fields originate from gradients in the phase differences and relative densities of the components near domain walls and defects. 
In the three component Ginzburg-Landau model one can separate out the terms that do not depend on the superconducting $U(1)$ sector as follows:
\begin{eqnarray}
    F&=&    \frac{{\bf J}^2}{2e^2\varrho^2}
    +\frac{1}{2} {\bf B}^2\\
  	   & +&\sum_{i}\frac{1}{2}(\nabla|\psi_i|)^2
	 + a_i|\psi_i|^2+\frac{b_i}{2}|\psi_i|^4  
      \\ &+& \sum_{i<j}\frac{|\psi_i|^2|\psi_j|^2}{\varrho^2}\left(
\frac{[\nabla (\phi_i-\phi_j)]^2}{2}
      +\frac{\eta_{ij}\varrho^2\cos(\phi_i-\phi_j)}{|\psi_i||\psi_j|}\right)
      \label{SineGordonPart}.
\end{eqnarray}
Here, the first term is the kinteic energy of the supercurrent ${\bf J}\equiv \sum_{i} e\,{\rm Im}\left(\psi_i^*
D\psi_i  \right)$, and the second term is the standard magnetic-field energy density. The  total density is $\varrho^2=\sum_i|\psi_i|^2$. The terms in the second and third lines do not depend on the superconducting phase.

{  The expression for the $k$-component ($k=x,y,z$) of the magnetic field can be found by writing the standard expression for supercurrent that follows from the Ginzburg-Landau model, extracting the vector potential from it and taking a curl. The resulting terms can be expressed as follows \cite{Garaud2013}:}
\begin{eqnarray}
&B_k=\partial_l A_m-\partial_m A_l=-\epsilon_{lmk}\partial_l\left(\frac{J_m}{e^2|\Psi|^2}\right)
&-\frac{i\epsilon_{lmk}}{e |\Psi|^4} \left[
   |\Psi|^2\partial_l\Psi^\dagger\partial_m\Psi
   +\Psi^\dagger\partial_l\Psi\partial_m\Psi^\dagger\Psi
   \right] \,,
   \label{fields}
\end{eqnarray}
where $\Psi^\dagger=(\psi_1^*,\psi_2^*,\psi_3^*)$. The first term is the  contribution from standard superconducting  currents, while spontaneous fields originate from the second term, which has the form of $\CP^2$ Skyrmionic topological charge density  \cite{Garaud2013,Babaev2002a}. In a two-component case one gets a standard Skyrmionic topological charge density \cite{Babaev2002a}. Importantly, this second term depends only on the phase differences and relative densities of the condensates in the bands, i.e.\ physically it is associated with magnetic field induced by the counter-flow of components belonging to different bands {  (caused by phase-difference gradients). Note that phase-difference gradients alone do not induce magnetic fields since they do not lead to charge transfer in real space. However, magnetic fields do appear  if, in addition, there are  relative-density gradients caused by defects or thermal gradients. This is precisely why these fields are described by Skyrme-like terms that depend on both types of relative gradients.} In the superconducting state the magnetic field is partially screened by the standard London contribution, which is the first term in (\ref{fields}) \cite{garaud2014domain}. In the quartic metal state the superconducting part of the model is disordered, and the part corresponding to London screening is absent {  (cf.\ discussion at the level of the London model in \cite{babaev2004superconductor})}. On the other hand, the second term in (\ref{fields}) depends only on the gradients of the phase differences between the bands and not on the superconducting phase. Thus, that contribution is not directly affected by the superconducting phase transition, and vanishes only at the $Z_2$ transition that disorders the phase differences.

To write the effective model for the quartic metal one needs to separate variables and retain only degrees of freedom related to phase differences of the gaps, which are not affected by the proliferation of topological defects in the superconducting phase. The resulting effective $Z_2$ model of the quartic metal is:
\begin{eqnarray}
    F&=& \frac{1}{2} \left(\frac{i\epsilon_{lm}}{e^2|\Psi|^4} \left[
   |\Psi|^2\partial_l\Psi^\dagger\partial_m\Psi
   +\Psi^\dagger\partial_l\Psi\partial_m\Psi^\dagger\Psi
   \right]\right)^2
  \nonumber  \label{JPart} \\
\\ &+&\sum_{i}\frac{1}{2}(\nabla|\psi_i|)^2
	 + a_i|\psi_i|^2+\frac{b_i}{2}|\psi_i|^4  
 +\sum_{i<j}\frac{|\psi_i|^2|\psi_j|^2}{\varrho^2}\left(
\frac{[\nabla( \phi_i-\phi_j)]^2}{2}
      +\frac{\eta_{ij}\varrho^2\cos(\phi_i-\phi_j)}{|\psi_i||\psi_j|}\right).
      \label{SineGordonPart}
\end{eqnarray}
This model has the form of  a $\CP^2$ generalization of the Skyrme model \cite{skyrme1962unified} where the symmetry is explicitly broken down to $Z_2$.

{ 

\begin{figure}
	\centerline{
		\includegraphics[width=0.5\linewidth]{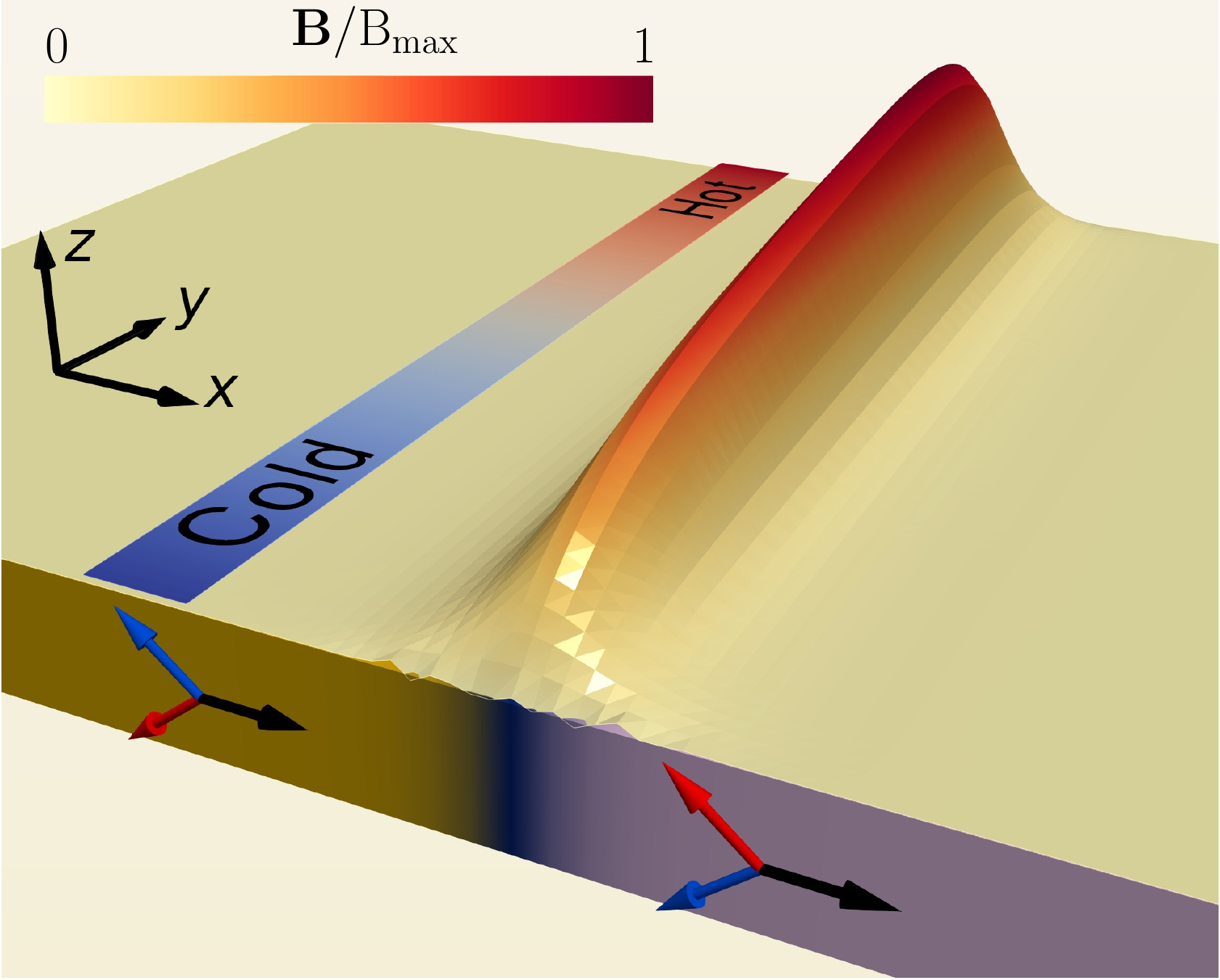}
 }
  \caption{\textbf{Spontaneous magnetic field} induced by a domain wall in the presence of thermal gradients in the BTRS quartic metal state. The surface elevation together with the colouring, represent the magnitude of the spontaneous magnetic field ${\bf B}$ \eqref{fields}, normalized to the maximal value $B_{\mathrm{max}}$. 
The domain wall itself is illustrated on the near face of the domain: the color scheme represents two different BTRS states, schematically depicted by arrows. The near and far faces are maintained at different temperatures, resulting in a thermal gradient along the domain wall.  Parameters are given in the text.}
  \label{FigDW}
\end{figure}

Figure \ref{FigDW} shows a numerical solution for the spontaneous magnetic field of a domain wall in the effective model \eqref{JPart}, \eqref{SineGordonPart}, in the presence of a thermal gradient along the domain wall. The complex fields $\Psi^\dagger=(\psi_1^*,\psi_2^*,\psi_3^*)$ are discretized within a finite-element formulation, and the effective free energy \eqref{JPart}, \eqref{SineGordonPart} is minimized by a nonlinear conjugate gradient algorithm. Detailed discussion of this numerical method can be found for example in the supplementary information of Ref.~\cite{garaud2016thermoelectric}. 
The domain wall interpolates in the $x$-direction between two different phase lockings. The thermal gradient is applied along the domain wall (in the $y$-direction), and is accounted for by having spatially dependent coefficients 
$a_i(y)=a_i^{(0)}g(y)$, where $g(y)$ grows linearly across the domain. For the results displayed in Fig.~\ref{FigDW}, the linear modulation is $g(0)=1$, and $g(L_y)=0.3$.
The potential parameters are $a_1^{(0)}=a_2^{(0)}=2a_3^{(0)}=-1$ and $b_i=1$. The parameter $e=0.25$, while the Josephson couplings are $\eta_{12} = \eta_{13} = -\eta_{23} = -2$.
For a given initial configuration, the nonlinear conjugate gradient algorithm converges to a domain-wall state with spontaneous magnetic field. This spontaneous field, which is induced by thermal gradients along the domain wall and is displayed in Figure \ref{FigDW}, has a shape quite similar to that of the magnetic field of a domain wall in the $s+\mathrm{i}s$ superconducting state \cite{silaev2015unconventional}. The  main difference is the   absence of a diamagnetic contribution, due to the absence of superconductivity. In the state considered here, as in the $s+\mathrm{i}s$ superconducting state \cite{garaud2014domain}, there are no magnetic signatures in the absence of thermal gradients or defects.}

The existence of a spontaneous Nernst effect requires a source of magnetic fields for the fluctuating disordered $U(1)$ sector. Such fields are caused by the term in (\ref{JPart}), which are not screened above the $U(1)$ transition. This is consistent with the increase of the spontaneous Nernst effect in the BTRS quartic metal phase above the superconducting phase transition. Note that in a multidomain phase the direction of magnetic field of domain walls will be alternating, but we expect non-zero net magnetic flux in a finite sample.

\subsection*{Monte-Carlo Calculations}

\subsubsection{Zero external field}
\begin{figure}
		\includegraphics[width=\textwidth]{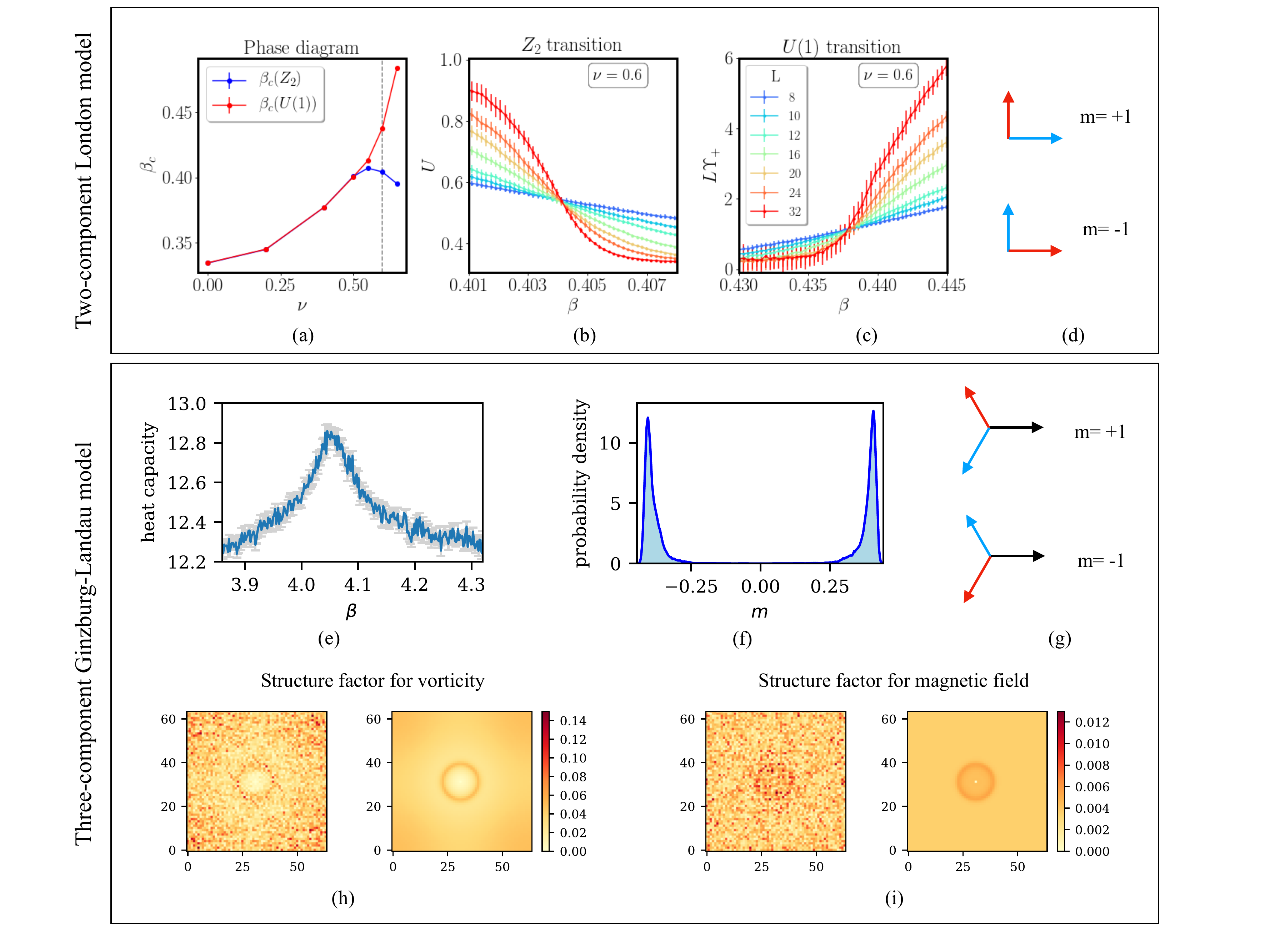}
	\caption{\textbf{Fluctuation-induced phases in multiband models.} Panels \textbf{(a)-(d)} show results for the two-component approximation of a three-band model in zero external magnetic field in the extreme type-II limit for various values of the mixed gradient coupling $\nu$.
	\textbf{(a)} Phase diagram.	The $U(1)$ and $Z_2$ transitions split apart for $\nu>0.5$, where the quartic metal phase emerges. The gray dashed line indicates the value $\nu=0.6$. \textbf{(b)} Binder cumulant $U$ at $\nu=0.6$ for different system sizes as function of the inverse temperature $\beta$. 
	\textbf{(c)} Helicity modulus for the phase sum, $\Upsilon_+$, at $\nu=0.6$ for different system sizes versus $\beta$.
	{\textbf{(d)} Illustrative example of the meaning of the two-component Ising order parameter $m$.}  Panels \textbf{(e) -(i)} Show results for {Three-component Ginzburg-Landau model.} \textbf{(e)} Heat capacity $L^{-3}\,\mathrm{d}\langle E\rangle/\mathrm{d}T$ versus $\beta$ for the system with applied field that we consider. 
	The heat capacity shows a signature of the $Z_2$ transition to a non-superconducting state associated with the breaking of time-reversal symmetry.
	{\textbf{(f)} }Histogram of the Ising order parameter $m$ for $\beta = 4.10$. For this inverse temperature the $Z_2$ symmetry is clearly broken. 
	{\textbf{(g)}} Illustration of the order parameter $m$ for the three-component case. 
	Structure factors for \textbf{(h)} the vorticity of $\psi_1$ and \textbf{(i)} the magnetic field, at the same inverse temperature $\beta = 4.10$ as for the above histogram. Snapshots are shown to the left and thermal averages to the right. In the presence of a vortex lattice, the structure factors will have pronounced peaks. The absence of such peaks indicates that the system is in a resistive vortex-liquid state which breaks $Z_2$ symmetry due to nontrivial phase locking. (We remove the trivial zero-wave-vector components of the structure factors for clarity, and normalize the remaining components to the zero-wave-vector component.)}
	\label{FigED3}
\end{figure}

We begin by considering the minimal three-band Ginzburg-Landau model in zero external field. The Monte-Carlo calculations of Ginzburg-Landau models were performed using the Metropolis-Hastings algorithm with local updates of the matter fields and the vector potential. No gauge fixing was used (all observables are gauge-invariant quantities). We also used parallel-tempering swaps between systems with neighboring temperatures; typically one set of swaps was proposed every $16$ or $32$ sweeps. The sizes of the local updates were adjusted during the equilibration in order to make the acceptance probability for each type of local update $50\, \%$ (except that phases were in some cases updated by simply picking a new value independent of the old value). Also, the simulated temperatures were adjusted within a fixed interval in order to make the acceptance ratios for parallel-tempering swaps equal for all pairs of neighboring temperatures.

First, we assess the limit of zero external field. In this limit, we begin by considering the parameter values that are most favorable to the occurrence of the quartic $Z_2$ metal phase in the class of latticized  models given by (\ref{contf}). That is, we consider the conditions where the spontaneous breaking of the $Z_2$ symmetry is the strongest in this class of models. The $Z_2$ symmetry is restored when domain walls proliferate, and the $U(1)$ symmetry is restored when vortices proliferate. Thus, in order to constrain the class of models we consider, we consider the case when domain walls are as energetically expensive as possible, relative to vortices (see the derivation in the supplementary information). Assuming that the superconductor is type-II near the superconducting phase transition, our simulation results on the simplest model indicate that, for the lattice sizes we consider, the stiffness of the $Z_2$ order parameter is not strong enough in the minimal model (\ref{contf}) to produce a fluctuation-induced BTRS quartic metal phase in exactly zero external magnetic field. 
However going to larger lattice sizes in this type of problems often shows that there exists a small four-fermionic phase rather than a single first order phase transition \cite{Herland2013}.

However, our experimental results strongly suggest that a substrantial BTRS quartic metal phase is present even in the limit of zero magnetic field. This suggests that the cost of domain-wall excitations relative to vortex excitations is underestimated in our 
basic model.
Our results suggest that in this material the situation may be more complex than what can be described by the simplest Ginzburg-Landau expansion corrected by fluctuations. This is consistent with the experimental observation that the pair-formation crossover takes place at twice as high temperature as the symmetry-breaking transitions. This suggests that London-based model \cite{Bojesen2014a} might be more appropriate.

{ Next we show that phenomenologically taking into account  mixed gradient terms gives rise to a $Z_2$ phase in zero external field even in the the extreme type-II limit. 
Mixed gradient terms are generically present in multicomponent systems and originate for example from Fermi-liquid corrections or strong correlations as shown in various physical contexts~\cite{leggett1975theoretical, Sjoberg:76, kuklov2004superfluid,Kuklov2006, Svistunov2015, Sellin.Babaev:18}. This is one of the terms that changes the relative energy cost of domain walls. 
In this investigation, we neglect density-field fluctuations setting  $|\psi_{1,2}|=\rho_{1,2}$, and use an approximated model, where the three-component order parameter is projected to a two-component one \cite{Garaud2017}. Furthermore, we consider the case of infinite magnetic field penetration length (i.e.\ the extreme type-II limit). The latter two approximations significantly underestimate  the size of the $Z_2$ phase. However, as emphasised in the above, here our} goal is to show that the $Z_2$ phase appears in zero field even in this approximation when one takes into account the mixed gradient terms. The resulting functional reads:

\begin{equation}
    f=  \sum_{i=1}^2 \frac{\rho_i}{2} \left( \mathbf{\nabla} \phi_i \right)^2 - \nu \left(\mathbf{\nabla} \phi_1 \cdot \mathbf{\nabla}\phi_2 \right) + \eta_2 \cos[2(\phi_1 -\phi_2)]. 
\label{2c_london}
\end{equation}

The higher-order Josephson interaction term~\cite{Maiti2013,garaud2016thermoelectric}, with coupling constant $\eta_2$, locally couples the two superfluid components. As for the three-component case, we can define a $Z_2$ Ising order parameter $m$ relative to the two possible phase differences: $m=\pm 1$ accordingly with $\Delta \phi_{1,2}= \pm \pi/2$ [see Fig.~\ref{FigED3}-(d) for an illustrative example]. We find that, when $\nu$ is sufficiently large, the quartic metal state appears even in the limiting case $\lambda \to \infty$, as shown in the upper panel of  Fig.~\ref{FigED3}. The $Z_2$ inverse critical temperature $\beta_c(Z_2)$ in Fig.~\ref{FigED3}-(a) has been determined from the finite-size crossings of the Binder cumulant $U$ [see Fig.~\ref{FigED3}-(b)], while the $\mathrm{U}(1)$ inverse critical temperature $\beta_c(\mathrm{U}(1))$ has been determined from the finite-size crossings of $L \Upsilon_+$ [see Fig.~\ref{FigED3}-(c)], where $\Upsilon_+$ is the helicity modulus for the phase sum. More details can be found in the supplementary information.

\subsubsection{Finite external field}

While our simulations of the lattice version of the minimal model (\ref{contf}) underestimate the domain-wall energy and thus the presence of the quartic phase in zero field, nonetheless our calculations show that the model supports the presence of the BTRS quartic 
metal phase in non-zero magnetic field. These calculations confirm that there is a specific-heat signature at the $Z_2$ phase transition inside the vortex liquid state (Fig.~\ref{FigED3}-(e)).  Our simulations indicate that the size of the anomaly is expected to be small compared to dominant mean-field contribution in the specific heat. The entropy change at the $Z_2$ transition is related to a disordering of the interband phase differences, which is small compared to entropy change associated with the formation of Cooper pairs. Also, the experimental results indicate a second-order phase transition since no signature of hysteretic behaviour was found in the transport and thermodynamic properties. 
The presence of a quartic metal phase for the model we consider is illustrated by Figs.~\ref{Fig1} and \ref{FigED3}. In Fig.~\ref{FigED3}-(f) we show a histogram of the Ising order parameter $m$ for the inverse temperature $\beta = 4.10$. This histogram clearly shows that the $Z_2$ symmetry associated with interband phase differences is spontaneously broken  (note that the external magnetic field does not explicitly break the $Z_2$ symmetry associated with the interband phase differences). In Fig.~\ref{FigED3}-(h) and Fig.~\ref{FigED3}-(i) we show structure factors for the vorticity of $\psi_1$ (the three components are equivalent) and the magnetic field, at the same inverse temperature $\beta = 4.10$ as for the histogram. Both snapshots and thermal averages of the structure factors are shown. In the presence of a vortex lattice, the structure factors will have pronounced peaks. No such peaks can be seen, which demonstrates that the system is in a resistive vortex-liquid state where the superconducting phase is disordered, and yet there is a well defined Ising-type order parameter describing spontaneously broken time-reversal symmetry associated with phase differences between bands. This is in qualitative agreement with the experimental magnetic phase diagram shown in Figs.~\ref{Fig1} and \ref{FigED2}. The parameters used to produce Fig.~\ref{FigED3} are given in the supplementary information, both for the model \eqref{contf} and the model \eqref{2c_london}.

\subsection*{Samples}
Phase purity and crystalline quality of the plate-like Ba$_{1-x}$K$_x$Fe$_2$As$_2$ single crystals were examined by X-ray diffraction (XRD) and transmission electron microscopy (TEM). The $c$-axis lattice parameters were calculated from the XRD data using the Nelson-Riley function. The K doping level $x$ of the single crystals was determined using the relation between the $c$-axis lattice parameter and the K doping obtained in previous studies \cite{Kihou2016}. 
The selected single-phase samples had a mass $\sim 0.1 - 1$~mg with a thickness $\sim 10 - 50$~$\mu$m and a surface area of several mm$^2$. 

\subsection*{Experimental}
DC susceptibility measurements were performed using a commercial superconducting quantum interference device (SQUID) magnetometer from Quantum Design. The measurements of the specific heat using the thermal relaxation method, electrical transport, and AC susceptibility shown in Figs.~\ref{Fig2},and~\ref{ED_0p8_PD} were measured in a Quantum Design physical property measurement system (PPMS). To boost the sensitivity at the specific heat measurements of small samples, the relative temperature increase was kept at 2 \%. This allowed to measure a broad dominant anomaly, but did not allow to resolve weak and relatively narrow fluctuation-induced anomalies. For high-resolution measurements we used the AC method (see below).   

\subsection*{Thermal transport measurements}
\begin{figure}
	\includegraphics[width=\columnwidth]{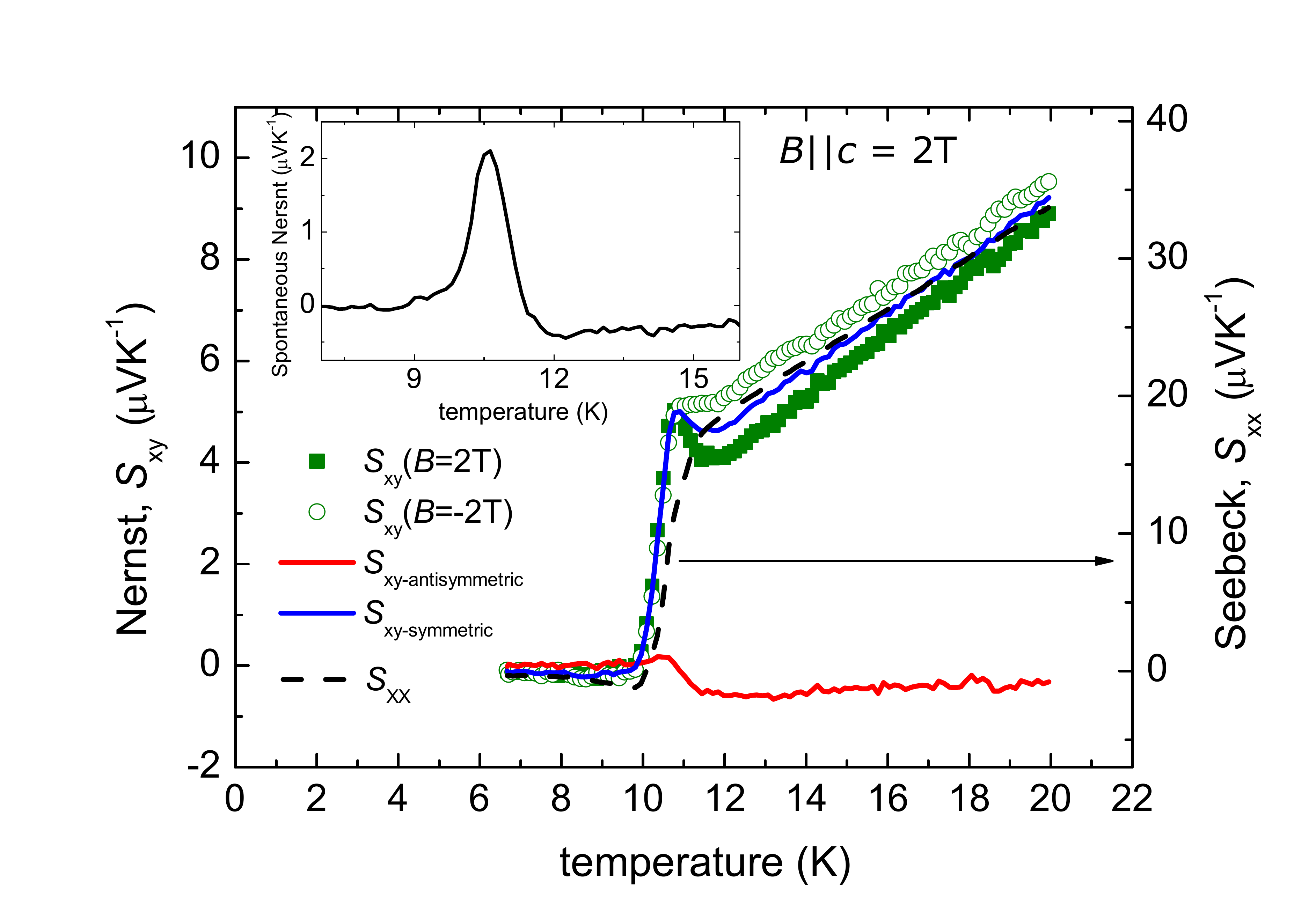}
	\caption{{\bf Extraction of the Spontaneous Nernst effect.} Temperature dependence of the Nernst (left axis) and Seebeck (right axis) signals for $x = 0.8$ and $B = 2$~T. The green full and empty symbols represent the Nernst signal for $B = 2$~T and $B = -2$~T, respectively. The red and blue lines represent the evaluated $B$-antisymmetric and $B$-symmetric parts of the Nernst signal, respectively. The black dashed line represents the Seebeck signal for $B = 2$~T. {\bf{Inset}}: Spontaneous Nernst effect at $B = 2$~T, extracted according to equation S1.}
	\label{FigED2_b}
\end{figure}

\begin{figure}
	\includegraphics[width= 8 cm]{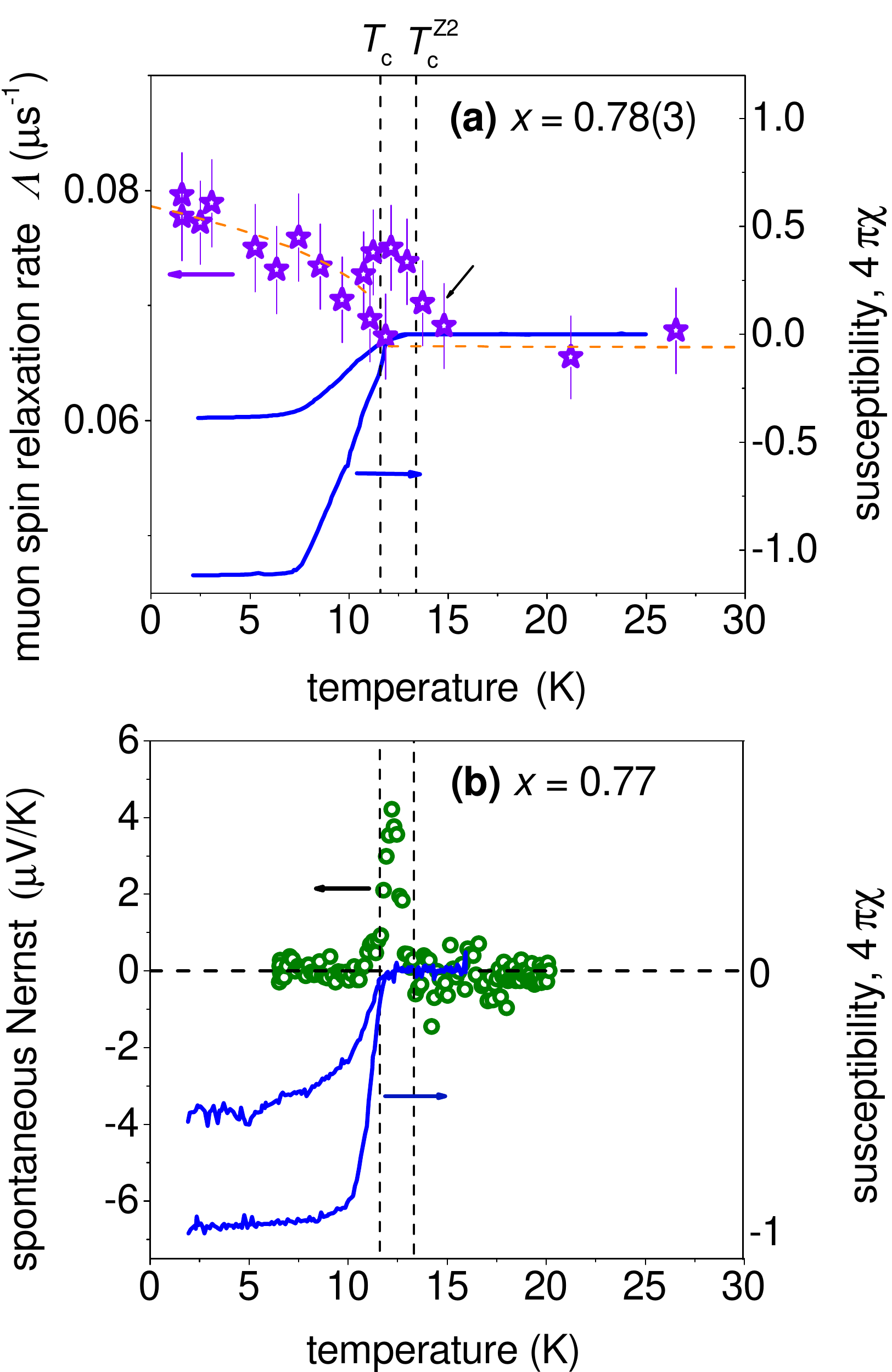}
	\caption{{\bf Comparison of different probes of BTRS state.}  {\bf (a)} Temperature dependence of the zero-field muon spin relaxation rate (left) shown together with the static magnetic susceptibility measured in $B\parallel ab = 0.5$~mT (right) for the stack of single crystals with $x$ = 0.78(3) \cite{Grinenko2020}. {\bf (b)} Temperature dependence of the spontaneous Nernst effect measured in zero magnetic field (left) shown together with the static magnetic susceptibility measured in $B\parallel ab = 0.5$~mT (right) for the sample with $x = 0.77$. The comparison between the $\mu$SR data and the spontaneous Nernst signal strongly suggests that the increase of the muon spin relaxation rate above $T_\mathrm{c}$ is not an artifact and that the origin of the spontaneous Nernst effect at $T_{\rm c}^{Z2}$ is spontaneous magnetic fields.}
	\label{muSR_ED}
\end{figure}

The Nernst- and Seebeck-effect measurements were performed using a home-made probe for transport properties inserted in an Oxford cryostat endowed with a $15$~T magnet. In order to create an in-plane thermal gradient on the bar-shaped samples, a resistive heater ($R = 2.7$~k$\Omega$) was connected on one side of the sample, while the other side was attached to a thermal mass. The temperature gradient was measured using a Chromel-Au-Chromel differential thermocouple, calibrated in magnetic field, attached to the sample with a thermal epoxy (Wakefield-Vette Delta Bond 152-KA). The Nernst and Seebeck signals were collected using two couples of electrodes (made of silver wires bonded to the sample with silver paint), aligned perpendicular to or along the thermal gradient direction, respectively. The magnetic field $B$ was applied in the out-of-plane direction. In order to separate the standard Nernst effect $S_{xy}$ from the spurious Seebeck component (caused by the eventual misalignment of the transverse contacts), the Nernst signal has been antisymmetrized by inverting the $B$ direction. The spontaneous Nernst signal, which is finite only in proximity to the superconducting transition, has been obtained by subtracting the Seebeck ($S_{xx}$) component from the $B$-symmetric part of the Nernst signal (Fig.~\ref{FigED2_b}), that is:
\begin{equation}
 \text{Spontaneous Nernst} = \frac{S_{xy}(B)+S_{xy}(-B)}{2}-kS_{xx},
  \label{sponer}
\end{equation}
where $k$ is a scaling factor obtained by dividing the Seebeck and the $B$-symmetric Nernst signals well above the superconducting transition. Of course, in zero field, equation S1 simply reads: Spontaneous Nernst = $S_{xy}-kS_{xx}$. The Seebeck and Nernst signals have been collected simultaneously in every measurement run in order to avoid any change in experimental conditions, which could in principle affect the extraction of the spontaneous Nernst effect. 

In the thermoelectric measurements for the samples with $x = 0.6$ and $x = 0.77$, the temperature difference $\Delta T_{\rm sample}$ across the sample (measured by the thermocouple) did not exceed $3\%$ of the measurement temperature $T$ fixed by the thermal mass. However, for $x = 1$ a considerably higher temperature gradient of about $\Delta T_{\rm sample} \approx 15\% T$ was applied due to a very low Nernst signal at low temperatures. Therefore, the noise level for $x = 1$ is considerably lower than for the other doping levels in Fig.~\ref{Fig4}.

Nernst effect  is a sensitive tool for detecting superconducting fluctuations \cite{Xu2000, Cyr2018}. In a Fermi-liquid state, the contribution to the 
Nernst effect is linear in temperature: $S_{\rm xy}/T \propto T$ \cite{Cyr2018}. As shown in Fig.~\ref{Fig4}, panels (d, e, f), this linear behavior is observed in Ba$_{\rm 1-x}$K$_{\rm x}$Fe$_2$As$_2$ in the normal state at high temperatures. The normal-state Nernst effect is small and positive for the samples with $x = 0.6$ and $x = 1$, but it is large and negative for $x = 0.77$. The change of the sign and magnitude of the Nernst effect indicates a significant change in normal-state properties triggered by the Lifshitz transition at $x \sim 0.60$ \cite{Grinenko2018, Hodovanets2014}.  The experimental data deviate from the linear dependence a few kelvin above $T_{\rm c}$ for the sample with $x = 0.6$, but for the sample with $x = 0.77$, which has a BTRS state, the deviation is observed at $T_{\rm SF}^{\rm high} \sim 2T_{\rm c}$. The temperature $T_{\rm SF}^{\rm high}$ is field dependent and scales with $T_{\rm c}$ (Fig.~\ref{Fig1}c), indicating a direct relationship with superconductivity. This allows us to associate $T_{\rm SF}^{\rm high}$ with a characteristic crossover temperature where significant superconducting fluctuations set in. At $T_{\rm SF}^{\rm high}$ we also found a deviation from the normal state $T^2$ behavior in the temperature dependence of the electrical resistivity (Fig.~\ref{Fig4}h). The same behaviour is observed also for the samples with $x = 0.8$ and $x = 0.81$ (Figs.~\ref{Fig2} and~\ref{ED_SH_S14}, respectively). 
However, the temperature range with pronounced superconducting fluctuations is much narrower for the doping levels away from $x \sim 0.8$. For the samples with $x = 0.6$ ($T_{\rm c} = 24$K in zero field), a deviation from the normal-state behavior is observed below 1.2$T_{\rm c}$ only, in agreement with the Nernst-effect data [panels (d, and g)]. (We note that a specific situation occurs at doping $x \sim 0.60$, for which  
the resistivity in a broad temperature range { $\rho\propto T^{1.66}$}. Such behavior is not unexpected at this special point due to the proximity to the Lifshitz transition \cite{Barber2018}.) In the case of KFe$_2$As$_2$ ($x = 1$, and $T_{\rm c} = 3.4$K in zero field), in-field Nernst effect measurements in the superconducting state are challenging due to the low upper critical field. However, combining the Nernst-effect data (panel f) with the resistivity data (panel i) we conclude that superconducting fluctuations do not extend above $T_{\rm SF}^{\rm high} \sim 1.5T_{\rm c}$.   

\begin{figure}
	\includegraphics[width=\columnwidth]{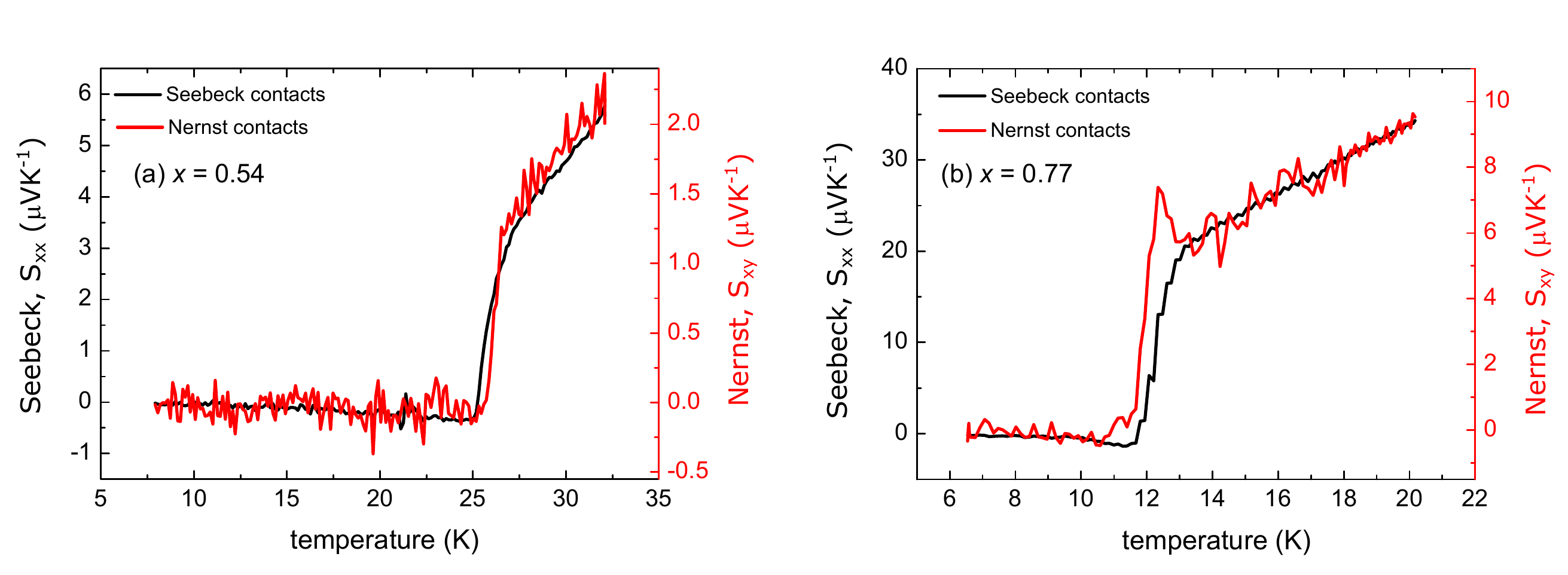}
	\caption{{\bf Raw thermoelectric data.} Temperature dependence of the zero-field voltage measured with Seebeck (left axis) and Nernst (right axis) contacts when a temperature gradient is applied for {\bf (a)} $x = 0.54$ and {\bf (b)} $x = 0.77$. A clear difference between the signals is seen in panel (b) only.}
	\label{FigED2}
\end{figure}

\begin{figure}
	\includegraphics[width=\columnwidth]{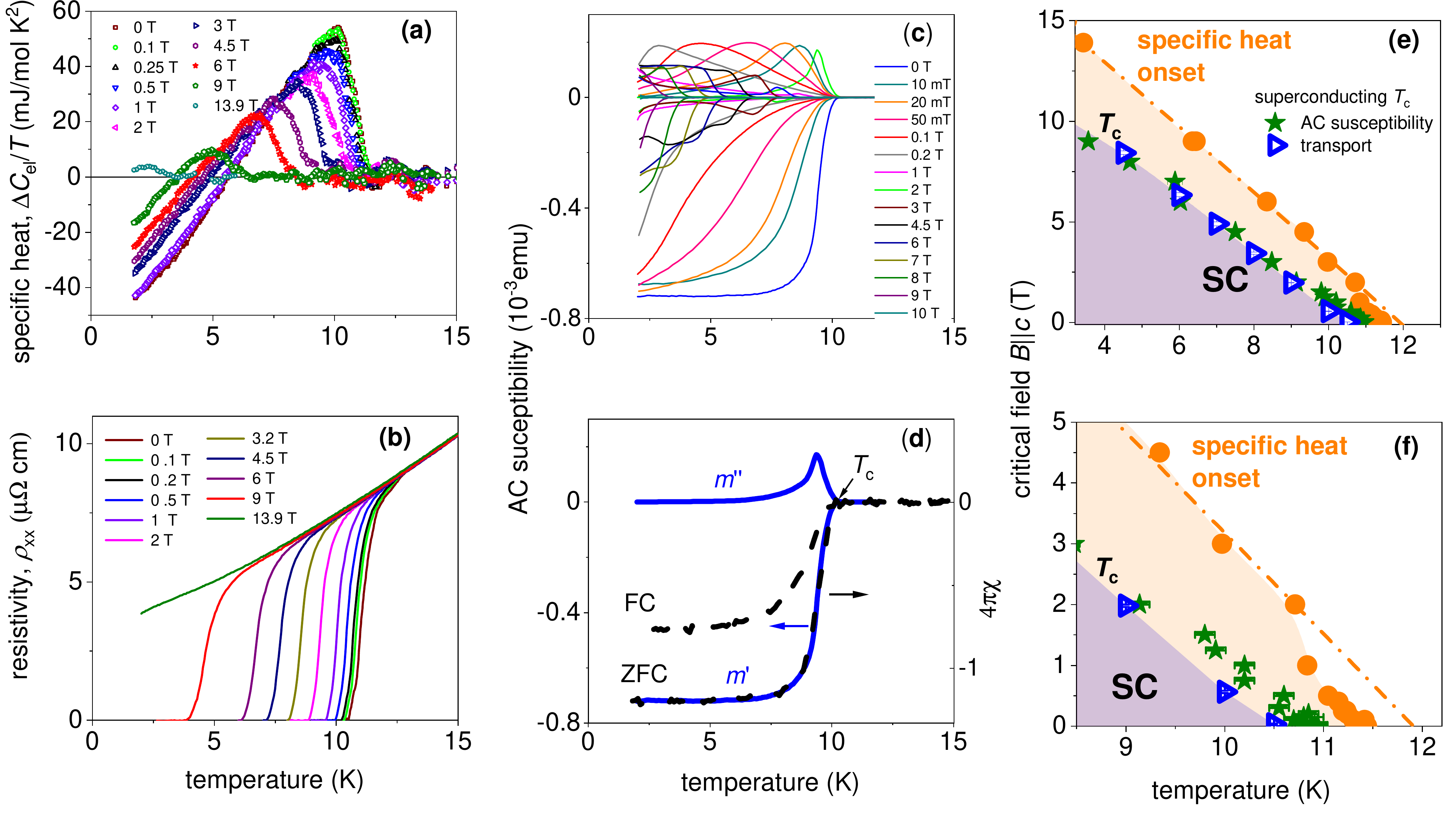}
	\caption{{\bf In-field properties of the sample with $x$ = 0.8}	{\bf (a)} Temperature dependence of the specific heat  measured by the relaxation method with the relative temperature increase of 2\%. {\bf (b)} Temperature dependence of the longitudinal electrical resistivity measured with the AC current amplitude $1$~A/cm$^2$ and the frequency $173$~Hz. {\bf (c)} Temperature dependence of the AC susceptibility measured with AC excitation field $B = 0.3$~mT and $f = 777$~Hz in different DC fields applied along the crystallographic $c$-axis. {\bf (d)} Temperature dependence of the AC susceptibility measured with AC excitation field $B = 0.3$~mT and $f = 777$~Hz in zero DC field (left axis), and the DC susceptibility in $B = 0.5$~mT applied in the $ab$-plane (right axis). {\bf (e)} Experimental magnetic-field phase diagram for the superconducting phase transition defined by zero resistance from the data in panel (b), and AC susceptibility from the data in panel (c), and onset  temperature of the dominant specific heat anomaly shown in panel [Note that the resolution here is not sufficient to identify $T_c^{Z2}$] (a). The relative splitting between the temperatures is increased with magnetic field. {\bf (f)} The low-field region of the same phase diagram as in panel (e).}
	\label{ED_0p8_PD}
\end{figure}

\begin{figure}
	\includegraphics[width=\columnwidth]{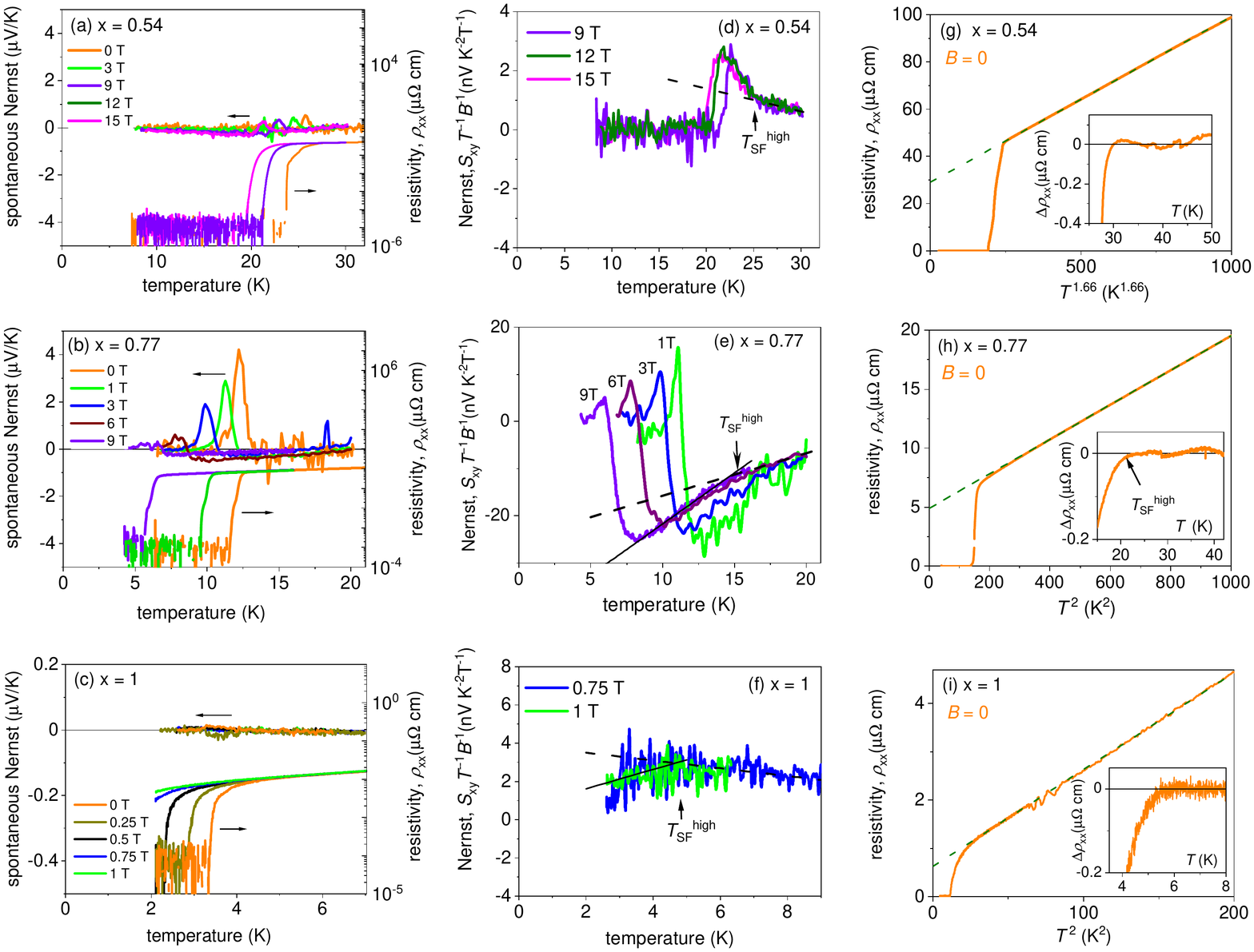}
	\caption{{\bf Spontaneous Nernst effect and superconducting fluctuations.} {\bf (a-c)} Temperature dependence of the spontaneous Nernst effect (left) and electrical resistivity (right) for two Ba$_{\rm 1-x}$K$_{\rm x}$Fe$_2$As$_2$ samples with doping levels $x = 0.54$ and $x= 1$ (without quartic phase), compared with the very different behavior of a sample with $x = 0.77$ that has a quartic phase. 
	A strong spontaneous Nernst signal was observed for $x = 0.77$, which is clearly seen in the raw data shown in Fig.~\ref{FigED2}b. 
	{\bf (d-f)} Temperature dependence of the conventional (odd in magnetic field) Nernst effect ($S_{\rm xy}$) for the same samples. Dashed lines show a linear behavior in temperature for the quasiparticle contribution in the normal state, $S_{\rm xy}/T$. For the sample with $x = 0.54$, field dependence in the Nernst signal appears a few kelvin above $T_{\rm c}$. For the sample with $x = 0.77$, the Nernst effect has a complex behaviour: it becomes field dependent at $T_{\rm SF}^{\rm high} \sim 2T_{\rm c}$, has a minimum roughly at $T_{\rm c}^{\rm Z2}$, a maximum at $T_{\rm c}$, and goes to zero at $T_{\rm SF}^{\rm low}$ (see also Fig.~\ref{Fig3}). $T_{\rm SF}^{\rm high}$ is the onset temperature of detectable superconducting fluctuations, and $T_{\rm SF}^{\rm low}$ is the temperature were fluctuations become undetectable. {\bf (g-i)} Temperature dependence of the longitudinal electrical resistivity in zero field. Solid curves are the experimental data and dashed lines are fits in the normal state (for explanation see the main text). Insets show the temperature dependence of the difference between the fit curves and the experimental data. The resistivity deviates from the normal-state behaviour at $T_{\rm SF}^{\rm high}$.}
	\label{Fig4}
\end{figure}

\begin{figure}
	\includegraphics[width=\columnwidth]{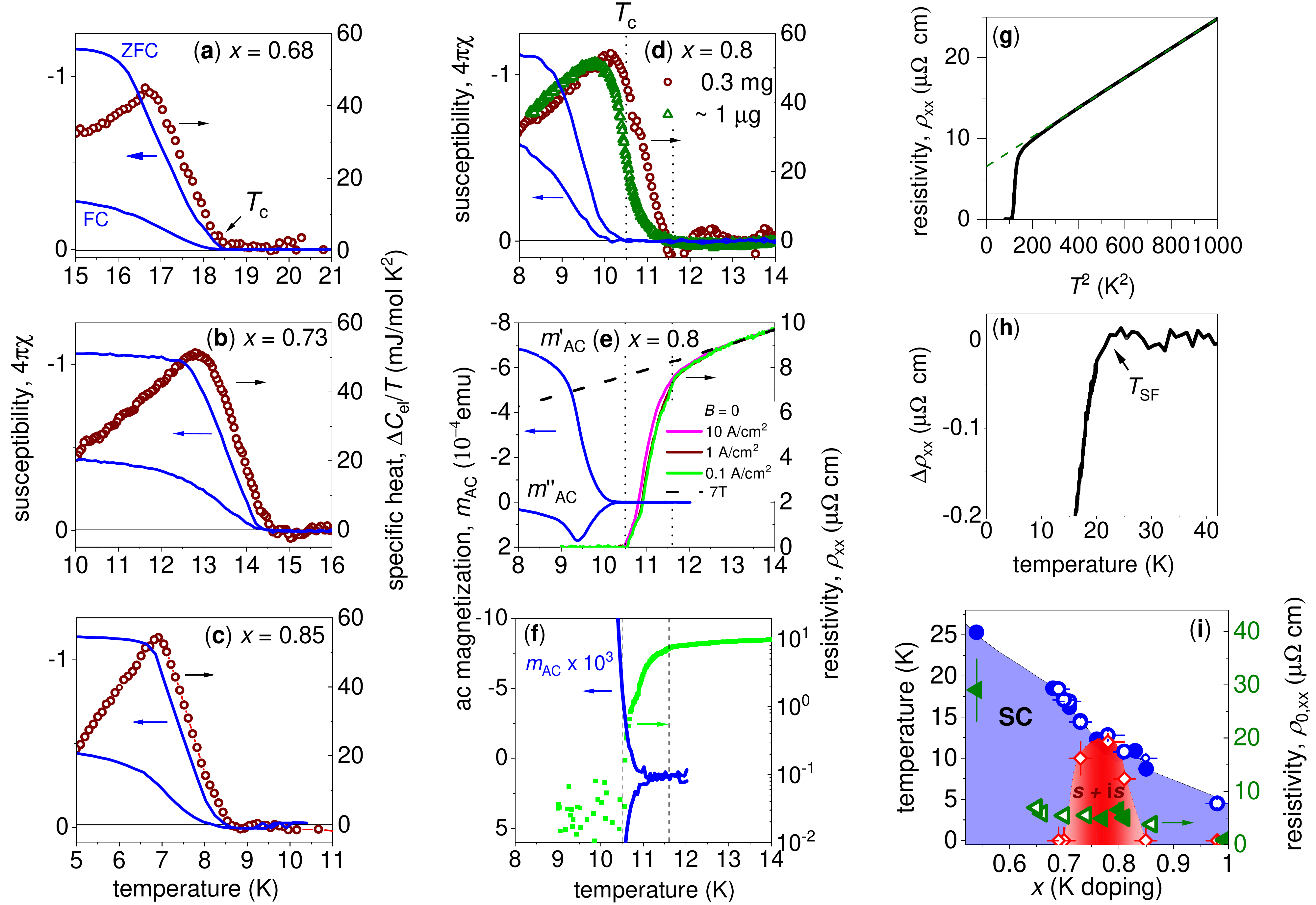}
	\caption{
	{\bf Ba$_{1-x}$K$_x$Fe$_2$As$_2$ at different doping levels $x$. (a-d)} 
	Temperature dependence of the magnetic susceptibility (left axis) measured in $B\parallel ab = 0.5$~mT applied after cooling in zero field (ZFC) with consequent measurement wile cooling in the same field (FC) and the zero-field specific heat (right axis) measured by the relaxation method with the relative temperature increase of 2\% 
	for samples with different K doping level $x$ and mass $m_{\rm s} \sim 1$~mg.  
	Two specific heat curves, shown in panel {\bf (d)}, were measured using different techniques: the sample with mass $m_{\rm s} \sim 1$~$\mu$g measured by microcalorimetry was cut from the larger sample with $m_{\rm s} = 0.3$~mg measured by the relaxation technique. {\bf (e)} Temperature dependence of the ac magnetization (left axis) measured in $B_{\rm ac}\parallel c = 0.3$~mT and at $f_{\rm ac} = 777$~Hz and electrical resistivity (right axis) measured in zero magnetic field and at different applied currents. The dashed curve is the electrical resistivity at $B\parallel c = 7$ T. The data is obtained on the same sample with $m_{\rm s} = 0.3$ mg as in {\bf (c)}.
	{\bf (f)} Temperature dependence of the ac susceptibility magnified by a factor of $10^3$ (left axis) and electrical resistivity in a log-scale (right axis) measured in zero magnetic field. Both the resistivity and the susceptibility give the same $T_{\rm c}$,  putting a constraint on  inhomogeneity.
	{\bf (g)} Temperature dependence of the resistivity in zero magnetic field (solid line) and a $T^2$-fit (dotted line). {\bf (h)} Temperature dependence of the difference between a $T^2$-fit and the data shown in panel (g). The observed behaviour is very similar to that for the sample with $x = 0.77$ shown in Fig.~\ref{Fig4}h.
	{\bf (i)}Experimental phase diagram (left axis): the superconducting transition is
	  extracted from  the DC susceptibility data (closed symbols) and taken from Ref.~\cite{Grinenko2018} (open symbols); the $s$ +i$s$ dome is taken from Ref.~\cite{Grinenko2018}.  Green triangles (right
	  axis) show the doping dependence of the residual resistivity $\rho_{xx,0}$ of the
	  Ba$_{\rm 1-x}$K$_{\rm x}$Fe$_2$As$_2$ single crystals.} 
	\label{Fig2}
\end{figure}

\begin{figure}
	\includegraphics[width=\columnwidth]{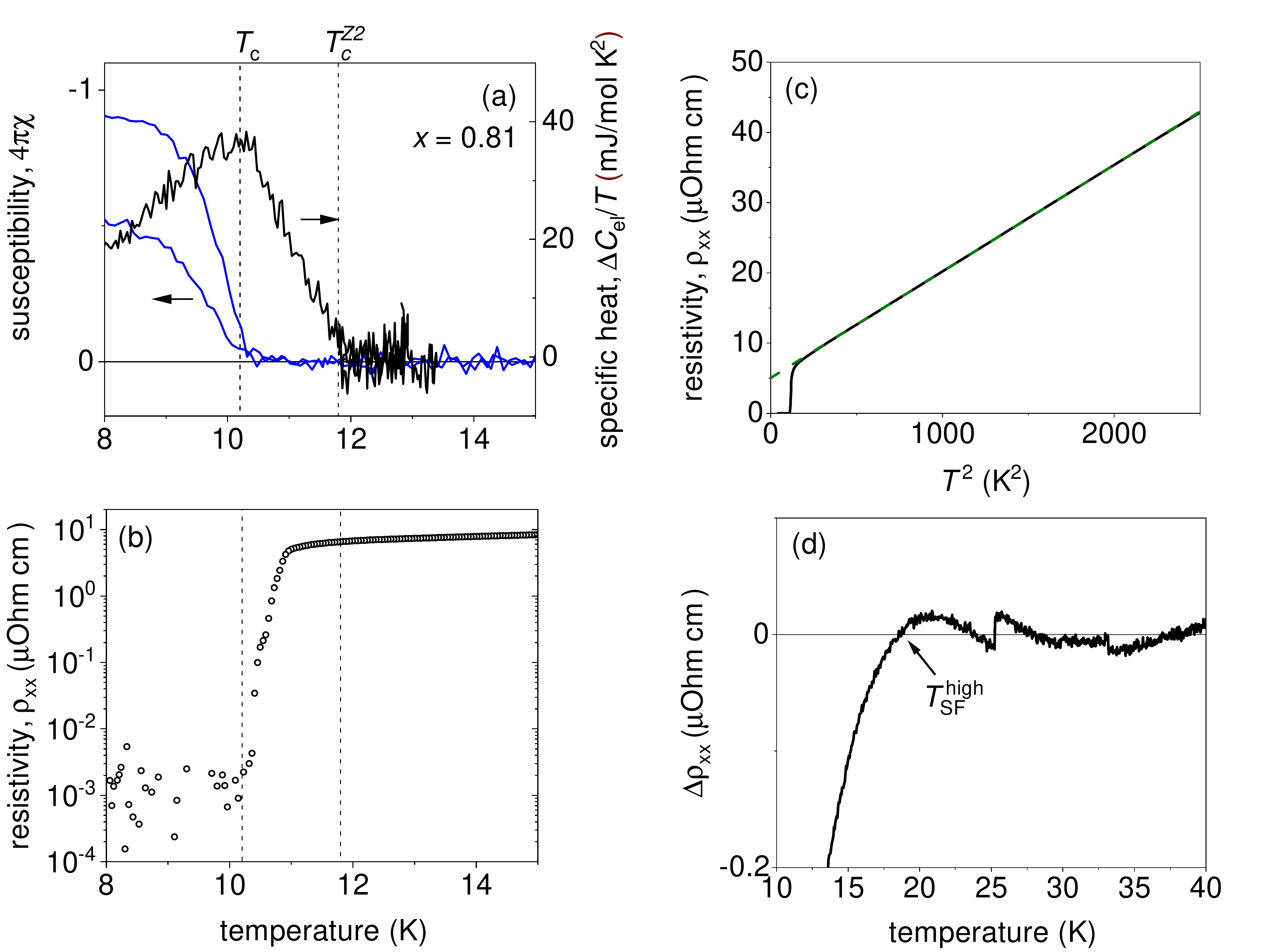}
	\caption{{\bf Sample with $x = 0.81$, $m_{\rm s} = 0.1~{\rm mg}$.} {\bf (a)} Magnetic susceptibility in $B\parallel ab = 0.5$~mT (left) and zero-field specific heat (right) measured by the AC technique. {\bf (b)} Temperature dependence of the electrical resistivity in zero magnetic field. Both the resistivity and the susceptibility give the same $T_{\rm c}$, indicating that the sample is  significantly homogeneous. An anomaly in the specific heat is observed above $T_{\rm c}$. {\bf (c)} Temperature dependence of the resistivity in zero magnetic field. {\bf (d)} Temperature dependence of the difference between a $T^2$ fit and the data. The observed $T^2$ behaviour is very similar to the sample with $x = 0.77$ shown in Fig.~\ref{Fig4}h in the main text.}
	\label{ED_SH_S14}
\end{figure}

\subsection*{AC calorimetry}
The heat capacity was measured in a purpose-built calorimeter using the thermal relaxation method and ac calorimetry. For this, the sample was placed with Apiezon N grease onto a sapphire platform equipped with a Cernox thermometer and a resistive heater. After sample mounting, the setup was placed in a $^3$He sorb-pumped cryostat with a carefully calibrated RuO$_2$ thermometer. For thermal relaxation experiments, the heater was powered with a constant output until thermal equilibrium was reached. The relative temperature increase was always kept at 0.5~$\%$. The heating and cooling branches were fitted with a single thermal relaxation time $\tau$.  For ac calorimetry, a sinusoidal voltage was applied to the heater –- in total 30 oscillations –- with a period of four or five seconds depending on the sample mass. The amplitude was always below 0.25~$\%$ of the absolute temperature resulting in a higher temperature resolution for the AC measurements. From a sinusoidal fit of the temperature profile, the heat capacity was calculated according to the work of Ref.~\cite{Sullivan1968}. During all experiments, the electrical resistance of the heater was measured in-situ by a four-point configuration where the current was applied via two wires and the voltage drop across the heater was measured via the other two wires.

\subsection*{Microcalorimetry}
The specific heat of the microgram-sized samples were performed using a fixed-phase ac steady-state method. The calorimeter cell consisted of a thin-film cermet thermometer, an offset (dc) heater, an ac heater, and a thermalization layer as described in Ref.~\cite{Willa2017}. Original samples were cleaved from all sides to suitable size and to obtain fresh surfaces. The measurements were performed in a Bluefors LD250 dilution refrigerator sitting at base temperature with local temperature control obtained through the use of the offset heater.

\subsection*{Disorder effect on the specific heat}
The specific-heat signatures of the fluctuation-induced $Z_2$ phase transition that we obtained have the form of small features on top of a broad single dominant anomaly. 
While on the one hand the broadness may originate in a washed out pairing crossover, the conventional and most common origin for the broadness of the mean-field (i.e. pairing-related) signature in the specific heat is the presence of intrinsic inhomogeneities. Some inhomogeneities should indeed be present in our case, since they are required to produce spontaneous magnetic field in an $s+is $ superconductor, and may contribute to the broadening. To exclude 
micron
-scale inhomogeneity effects, in Fig.~\ref{Fig2}c we also show the specific-heat data measured on a small fraction of the sample (118 x 147 x 7 $\mu$m$^2$) cleaved from all sides of the bulk single crystal. Both sets of specific-heat data show the same onset transition temperature, presenting further constraints against inhomogeities with these length scales. 

In some cases electronic inhomogeneities may result in the formation of incoherent Cooper pairs above $T_{\rm c}$ \cite{Sacepe2011}. However, our most unusual observation is the lack of any detectable diamagnetic response 
beyond the percolation threshold for 3D systems \cite{Grinenko2006}. This observation is inconsistent with inhomogeneous nucleation of superconductivity, including nanoscale separation with a droplet size well below the superconducting penetration depth.  
 In addition, such inhomogeneities would increase the electrical resistivity at low temperatures. To check this scenario, we performed systematic measurements of the electrical resistivity of samples with various doping levels. We observed that the residual resistivity value decreases with increased K doping, without there being any anomalous behaviour for the doping level with the $Z_2$ phase (Fig.~\ref{Fig2}i). This observed lack of increase in electron scattering rates points against the scenario where electronic inhomogeneities are stronger at $x \sim 0.8$. 
Furthermore, to the best  of our knowledge, there are no theoretical models or experimental examples where a spontaneous Nernst effect would arise from inhomogeneities in a non-BTRS superconductor.
If time-reversal symmetry is spontaneously broken by the formation of Cooper pairs,
the existence of the BTRS quartic metal state does not require a homogeneous sample, and may be realized also as a Josephson junction array made of BTRS superconductors.

\section{Acknowledgments}

\begin{acknowledgments}
The work was supported by DFG (GR 4667, CA 1931/1-1 (F.C.), GRK 1621, SFB 1143 (project-id: 247310070), and the Würzburg-Dresden Cluster of Excellence on Complexity and Topology in Quantum Matter–$ct.qmat$ (EXC 2147, Project ID 390858490) and the Swedish Research Council Grants No.\ 642-2013-7837, 2016-06122, 2016-04516, 2018-03659 and by the G\"oran Gustafsson Foundation for Research in Natural Sciences and Medicine. This work was performed in part at the Aspen Center for Physics, which is supported by National Science Foundation grant PHY-1607611. The simulations were performed on resources provided by the Swedish National Infrastructure for Computing (SNIC) at the National Supercomputer Center at Link\"oping, Sweden. This work was supported by a Grant-in-Aid for Scientific Research on Innovative Areas "Quantum Liquid Crystals" (JP19H05823) from JSPS of Japan. This work has further been supported by the European Research Council (ERC) under the European Union’s Horizon 2020 research and innovation programme (Grant Agreement No. 647276-MARS-ERC-2014-CoG). Also, we acknowledge support of the HLD at HZDR, member of the European Magnetic Field Laboratory (EMFL). We acknowledge fruitful discussion with S.-L.~Drechsler, D.~Efremov, E. Herland, C.~Hicks, H.~Luetkens, Y.~Ovchinnikov and P.~Volkov. We are thankful to K.~Nenkov and C.~Klausnitzer for technical support.
\end{acknowledgments}

\renewcommand{\theequation}{S\arabic{equation}}
\renewcommand{\thefigure}{S\arabic{figure}}
\renewcommand{\thetable}{S\arabic{table}}
\setcounter{equation}{0}
\setcounter{figure}{0}
\setcounter{table}{0}

\pagebreak
\section{Supplementary information}
{\bf Here we present additional characterisation data for the samples with a quartic metal phase, and provide details of the calculations within Josephson-coupled three-component Ginzburg-Landau theory presented in the main text.}

\section{Experiment}

\subsection{Characterization of samples}

\begin{figure}
	\includegraphics[width=\columnwidth]{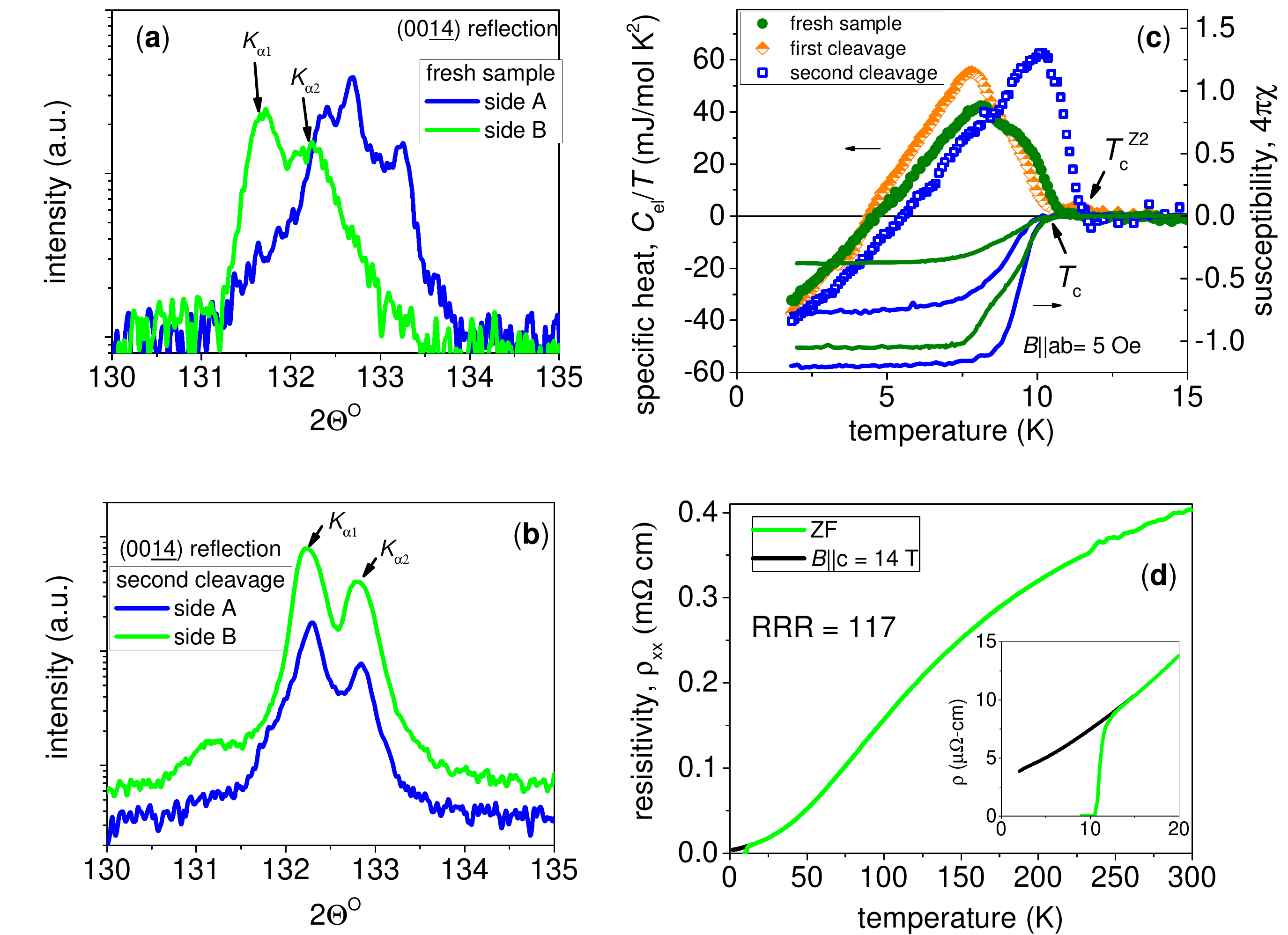}
	\caption{{\bf Preparation of the sample with $x = 0.8$.} {\bf (a)} X-ray scans measured from both sides of the plate-like fresh sample, just selected from a batch ($m_{\rm s} = 1.1$~mg). The position of high-angle reflections is not the same for the two sample sides corresponding to two slightly different doping levels. {\bf (b)} X-ray scans measured from both sides of the plate-like sample after the second cleavage ($m_{\rm s} = 0.3$~mg). The position of high-angle reflections is the same for two sample sides corresponding to the doping level $x = 0.8$. {\bf (c)} Temperature dependence of the specific heat (left axis) and the DC susceptibility (right axis) of the fresh sample and after the first and second cleavages. $T_\mathrm{c}$ is nearly unaffected by the cleavage but $T_{\rm c}^{\rm Z2}$ is clearly split off from $T_\mathrm{c}$ after the second cleavage. {\bf (d)} Temperature dependence of the electrical resistivity of the sample after the second cleavage measured over a broad temperature range. The inset shows the data at low temperatures.}
	\label{FigS1}
\end{figure}

It is a very challenging problem to grow homogeneous crystals in the doping range of interest. The usual problem is that several plate-like single crystals with a similar doping level are grown together face-to-face. To obtain a single-phase sample, we cleaved the crystals with sticky tape. The results of this procedure is demonstrated in Fig.~\ref{FigS1}. We found that the critical temperature $T_{\rm c}^{\rm Z2}$ is very sensitive to the sample conditions. For the single-phase sample with $x = 0.8$, the anomaly in the specific heat is observed at a higher temperature than for the initial double-phase sample. At the same time, $T_\mathrm{c}$ is nearly unaffected by the cleavage. The shift of $T_{\rm c}^{\rm Z2}$ is presumably caused by the strain which appears between two glued plate-like crystals (Fig.~\ref{FigS1}d). According to $\mu$SR experiments (see Fig.~5 in Ref.~\cite{Grinenko2018}), the BTRS dome in the phase diagram is very narrow in contrast to the nearly flat $T_\mathrm{c}$. Therefore, it appears that $T_{\rm c}^{\rm Z2}$ may be rather sensitive to the strain or that the interface has a strong gradient of local $T_{\rm c}^{\rm Z2}$, which is in line with the fact that $T_{\rm c}^{\rm Z2}$ has a strong doping dependence. The obtained single-phase samples had a high residual resistivity ratio $\mathrm{RRR} = \rho_{\rm 300K}/\rho_0 > 100$ as shown in Fig.~\ref{FigS1}c.

\begin{figure}
	\includegraphics[width=\columnwidth]{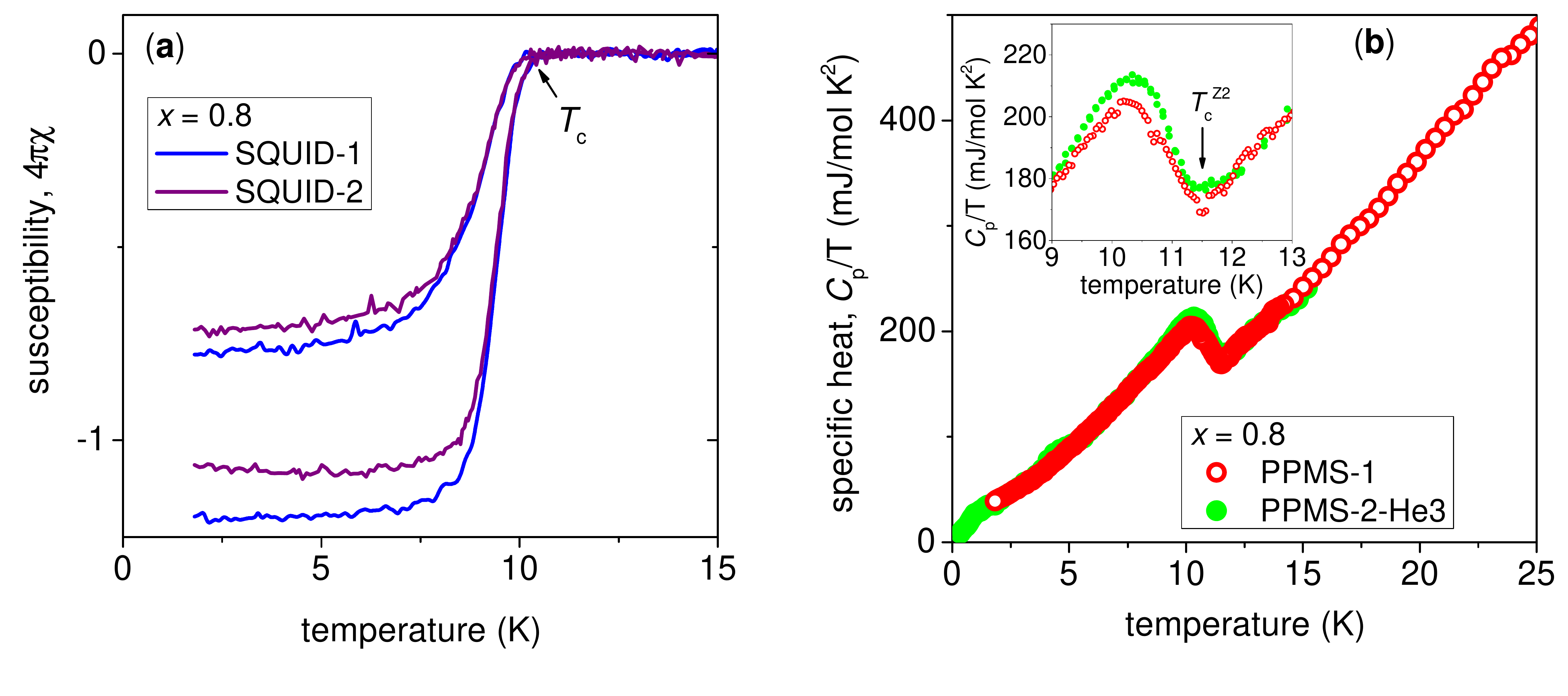}
	\caption{{\bf Reproducibility of the data for the sample with $x = 0.8$} (the data were collected after the second cleavage; see Fig.~\ref{FigS1}). {\bf (a)} Temperature dependence of the DC susceptibility measured in two different SQUID magnetometers. {\bf (b)} Temperature dependence of the specific heat measured in two different PPMS machines. The inset shows the data close to $T_{\rm c}^{\rm Z2}$.}
	\label{FigS2}
\end{figure}

We also took special care to exclude any experimental error in the measurements of the transition temperatures. To ensure this, we used different PPMS and SQUIDS devices in different institutes to make multiple measuments of the sample with $x = 0.8$ (the doping at which there is the largest splitting between $T_{\rm c}^{\rm Z2}$ and $T_\mathrm{c}$). As shown in Fig.~\ref{FigS2}, the results were essentially the same. Therefore, we exclude an experimental error in the temperature measurements on the scale of the observed effects.
\begin{figure}
	\includegraphics[width= 8 cm]{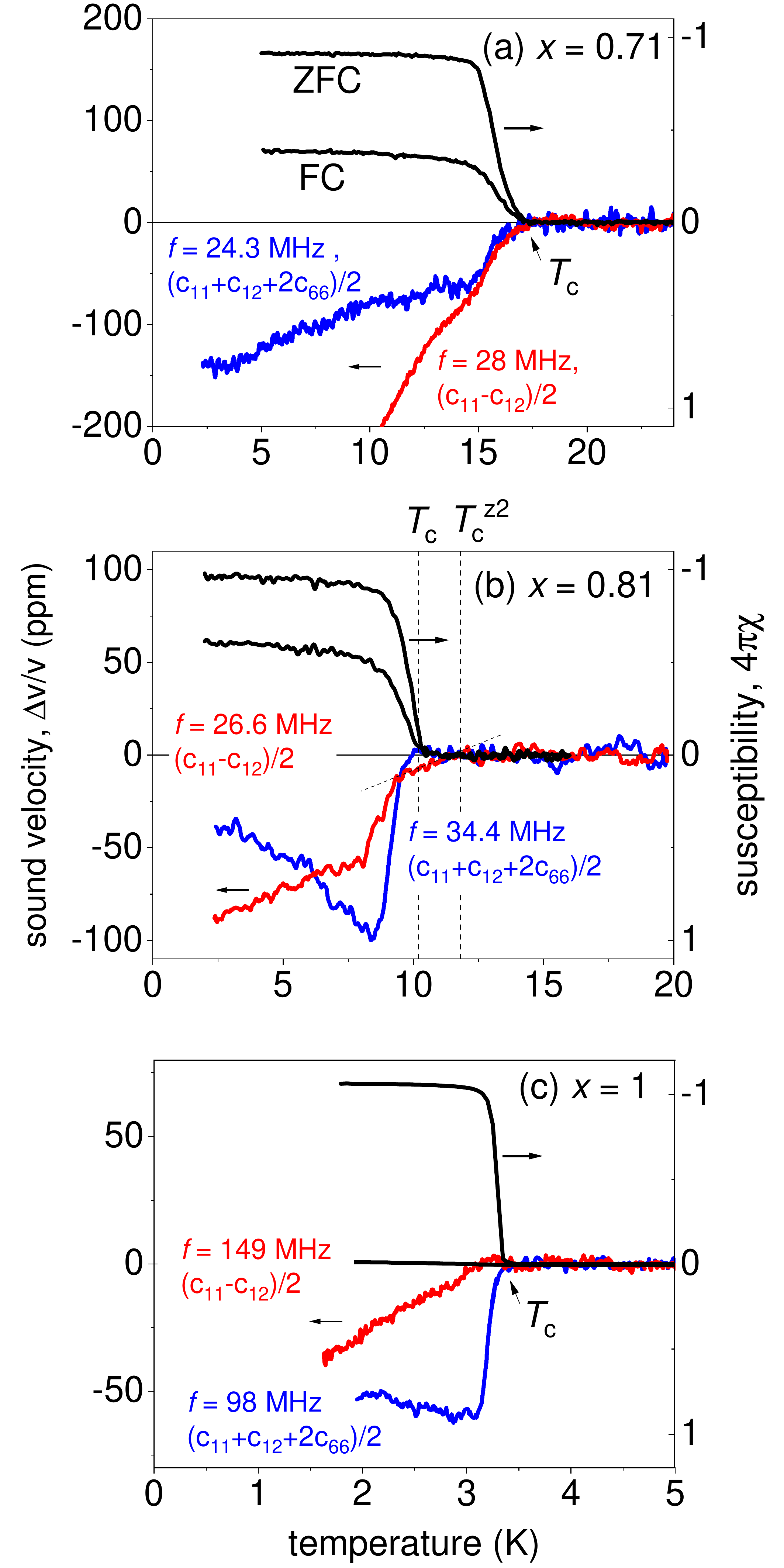}
	\caption{{\bf Ultrasound data.} Temperature dependence of the relative change of the sound velocity (left) for the longitudinal $(c_{11}+c_{12}+2c_{66})/2$ and transverse $(c_{11}-c_{12})/2$ acoustic modes with subtracted background (see Fig.~\ref{ultrsound_sup_0p81}) compared with temperature dependence of the magnetic susceptibility in $B\parallel ab = 0.5$~mT (right) measured after cooling in zero magnetic field (ZFC) and cooling in nonzero field (FC). The data for the samples with $x = 0.71$, $0.81$, and $1$ are shown in panels {\bf (a)}, {\bf (b)}, and {\bf (c)}, respectively. The sample with $x = 0.81$ exhibits a kink in the velocity of the transverse acoustic mode at a 
	position which agrees with the onset of the broad feature in the specific heat shown in Fig.~\ref{ED_SH_S14} for the same sample. The position of the kink agrees with the position of the $Z_2$ transition indicated by  the thermal transport experiments for the sample with  $x \sim 0.8$ (Fig.~\ref{Fig3} in the main text).
	}
	\label{ultrasound_chi_ED}
\end{figure}

\begin{figure}
	\includegraphics[width= 8 cm]{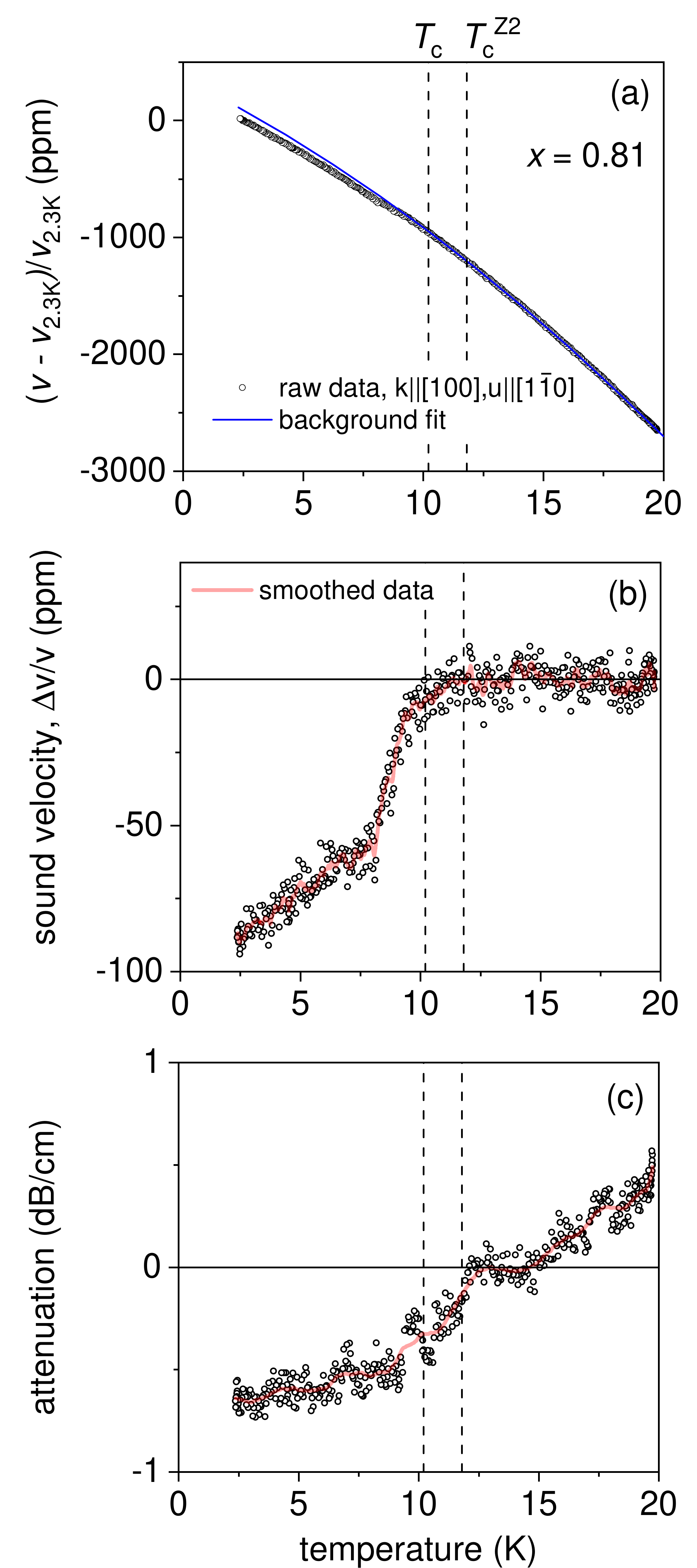}
	\caption{{\bf Ultrasound data analysis.} {\bf (a)} Temperature dependence of the relative change of the sound velocity of the transverse  $(c_{11}–c_{12})/2$ acoustic mode for the sample with $x = 0.81$. The measurements were done at $f = 26.6$ MHz using a transit acoustic signal (zero echo); the sample length was $L = 1.3$ mm and the thickness about $10~\mu$m. The blue solid line shows a background fit performed at $T > T_{\rm c}^{\rm Z2}$ using a quadratic polynomial function and extended to low temperatures. {\bf (b)} Temperature dependence of the relative change of the sound velocity obtained by subtracting the background fit from the raw data shown in panel (a). The red solid curve shows data smoothed by a simple averaging of the values of 5 points (these data are shown in Fig.~\ref{ultrasound_chi_ED}). {\bf (c)} Temperature dependence of the sound attenuation obtained at the same measurements. The data shows a clear anomaly close to $T_{\rm c}^{\rm Z2}$, supporting the conclusion that the kink in the temperature dependence of the sound velocity above $T_{\rm c}$ [panel (b)] is not an artefact of the measurements or data analysis.}
	\label{ultrsound_sup_0p81}
\end{figure}

\subsection*{Ultrasound measurements}

 Additional 
support for the existence of 
unconventional correlations above $T_{\rm c}$  is provided by ultrasound measurements (Fig.~\ref{ultrasound_chi_ED}).

The measurements were performed using a pulse-echo phase sensitive detection technique~\cite{Zherlitsyn_2014} in a gas-flow cryostat. A pair of piezoelectric LiNbO$_{3}$ resonance transducers were glued to parallel opposite (110) crystal surfaces in order to generate and detect acoustic waves. We used Z- and X-cut transducers (Boston Piezo-Optics Inc.) with fundamental frequencies close to 30~MHz for longitudinal $(c_{11} + c_{12} + 2c_{66}) / 2$ ($\boldsymbol{k} || \boldsymbol{u} || [110]$) and transverse $(c_{11} – c_{12}) / 2$ ($\boldsymbol{k} || [100], \boldsymbol{u} || [1\bar{1}0]$) acoustic modes, respectively. Here, $\boldsymbol{k}$ is the wavevector and $\boldsymbol{u}$ is the polarization of ultrasound waves.  Details of the data analysis are given in (Fig.~\ref{ultrsound_sup_0p81}).

Ultrasound velocity is related to the elastic constants, which are the second derivatives of the free energy with respect to the strains. Such thermodynamic derivatives typically exhibit a singular behavior at a  phase transition \cite{Luthi05}. For instance, at a superconducting phase transition, the sound velocity might show a kink or jump depending on the details of the superconducting order parameter and its coupling to the deformation tensor \cite{Nohara1995}.
The samples with doping levels away from  $x \sim 0.8$ show a 
conventional picture:  the anomaly in the sound velocity takes place at $T_{\rm c}$  with the concomitant onset of diamagnetic susceptibility. For the sample with $x = 0.81$ the situation is qualitatively different: there appears
  an additional singularity: a kink that coincides with
$T_{\rm c}^{\rm Z2}$   (Fig.~\ref{ultrasound_chi_ED}b). In addition, the sound attenuation 
shows an anomaly at $T_{\rm c}^{\rm Z2}$ as well (Fig.~\ref{ultrsound_sup_0p81}c).

\section{Theory}

\subsection{Minimal Ginzburg-Landau model of a three-band superconductor}

The experiments suggest that the quartic metal phase occurs only in a narrow
range of temperatures, and only close to the top of the dome of the $s+\mathrm{i}s$ phase.
Currently, there is insufficient knowledge on the microscopic physics of the material to derive a precise form of the Ginzburg-Landau theory. However, one can use the experiments to constrain the commonly used class of Ginzburg-Landau models with a minimal set of terms consistent with an $s+is$ state:
\begin{equation}
  f = \tfrac{1}{2}(\nabla \times \mathbf{A})^2
      + \sum_i \tfrac{1}{2} |(\nabla + \mathrm{i}e\mathbf{A}) \psi_i|^2
      + a_i |\psi_i|^2 + \frac{b_i}{2} |\psi_i|^4
      + \sum_{i<j} \eta_{ij}|\psi_i| |\psi_j| \cos(\phi_i-\phi_j).
  \label{contf2}
\end{equation}
First, let us assess whether this model has the quartic metal phase in zero magnetic field, assuming that the superconductor is type-II close to $T_c$.

Consider the question of what parameter values are most favorable for the occurrence of the quartic metal phase in the model (\ref{contf2}). First, it makes sense to make the three components symmetric, since differences between components would in general make one type of domain wall energetically preferred to the other two. Thus, we choose the parameters $a_i$, $b_i$ and $\eta_{ij}$ to be independent of $i$ and $j$, i.e.\ to be the same for all components and pairs of components. Second, increasing the strength of the Josephson coupling can hardly decrease the relative cost of domain walls, and thus we want the parameter $\eta$ to be large. Now, if one increases $\eta$ by a certain factor $k$, and then rescales the moduli of the matter fields, the vector potential, the spatial coordinates and the free-energy density itself correspondingly:
\begin{equation*}
  |\psi_i| \mapsto \sqrt{k}|\psi_i|,\ \ 
  \mathbf{A} \mapsto \sqrt{k}\mathbf{A},\ \ 
  \nabla \mapsto \sqrt{k}\nabla,\ \ 
  f \mapsto f/k^2,
\end{equation*}
then in terms of the rescaled quantities the free-energy density will be identical to the original one, except that $a$ is decreased by a factor of $k$ instead of $\eta$ being increased by the same factor. Thus, as far as the {\it relative} temperatures of the phase transitions are concerned, the limit of infinite $\eta$ is the same as the limit of zero $a$, which we choose to take. With $a$ set to zero there is, taking into account the above four freedoms to rescale, only one parameter in the continuum mean-field theory, which we take to be the electric charge $e$, whose role is to scale the penetration depth relative to the other length scales. The other remaining parameters in the expression for the free energy are set to $1$; thus, we have $a = 0$, $b = \eta = 1$. Now, including fluctuations we also have the inverse temperature $\beta$ as a parameter, and discretizing space we have the lattice constant $h$. Thus, the parameters we work with are the electric charge $e$ (which parameterizes the magnetic-field penetration length relative to density length scales), the inverse temperature $\beta$ and the lattice constant $h$.

\subsection{Length scales and normal modes}

Before describing the lattice model that we use for simulations, we describe another aspect of the continuum model, namely that of length scales and normal modes. In the simplest form of multicomponent Ginzburg-Landau theory, in which the matter fields interact only through their coupling to the vector potential, there will be one coherence length $\xi_i$ for each component. In other words, a small perturbation from the ground-state value of one of the matter-field amplitudes $|\psi_i|$ will not induce perturbations in any other degrees of freedom, and the amplitude will asymptotically recover its ground-state value exponentially in space with characteristic length scale $\xi_i$.

Now, consider the more general case of also having Josephson coupling between the $N$ components. If there is no time-reversal symmetry breaking, i.e.\ if the ground-state phase differences are all $0$ or $\pi$, then there will be $N$ density modes and $N-1$ phase (Leggett) modes, each with a corresponding length scale. (The final phase mode corresponds to a gauge transformation.) However, if time-reversal symmetry is broken, the normal modes will in general not be pure density or phase modes, instead being mixed phase-density modes \cite{Carlstrom2011b,Maiti2013,garaud2018properties}.

We have calculated ground states, length scales and normal modes for the model we consider (with the aforementioned parameters) in the way described in Ref.~\onlinecite{Carlstrom2011b}.  Namely, we have considered small perturbations in all degrees of freedom, linearized the free energy and solved the resulting eigenvalue and eigenvector problem to find the normal modes. We find that the normal modes are mixed, i.e.\ that the phase-difference modes are linearly coupled to the density modes. We find that the longest characteristic length scale associated with the matter fields has the value $1.20$ and the shortest the value $0.48$. There is a total density mode, for which all three densities vary in unison, with length scale $0.71$. The magnetic penetration depth reads:
\begin{equation}
  \lambda = \left( e \sqrt{\sum_i |\psi_i|^2} \right)^{-1},
\end{equation}
where each ground state density is $|\psi_i|^2 = 0.5$. The model is type-I when the penetration depth is the shortest length scale, i.e.\ when $e > 1.7$, and type-II when the penetration depth is the longest length scale, i.e.\ when $e < 0.68$.

\subsection{Monte Carlo simulation methods for Ginzburg-Landau model and observables}
In order to perform Monte Carlo simulations, we discretize the model (\ref{contf}) on a three-dimensional simple cubic lattice with $L^3$ sites and lattice constant $h$. The discretized model is given by the free-energy density
\begin{multline}
  f = \frac{1}{2h^2} \sum_{k<l} F_{kl}^2
      - \frac{1}{h^2} \sum_{i,k} |\psi_i(\mathbf{r})|
          |\psi_i(\mathbf{r}+\mathbf{k})| \cos \chi_{i,k}(\mathbf{r}) 
      + \sum_i \left( a_i + \frac{3}{h^2} \right) |\psi_i(\mathbf{r})|^2
          + \frac{b_i}{2} |\psi_i(\mathbf{r})|^4 \\
      + \sum_{i<j} \eta_{ij} |\psi_i(\mathbf{r})| |\psi_j(\mathbf{r})|
          \cos(\phi_i-\phi_j),
  \label{lattf}
\end{multline}
where
\begin{equation}
  F_{kl} = A_k(\mathbf{r}) + A_l(\mathbf{r} + \mathbf{k})
           - A_k(\mathbf{r} + \mathbf{l}) - A_l(\mathbf{r})
\end{equation}
is a lattice curl,
\begin{equation}
  \chi_{i,k}(\mathbf{r}) =
    \phi_i(\mathbf{r}+\mathbf{k}) - \phi_i(\mathbf{r}) + h e A_k(\mathbf{r})
\end{equation}
is a gauge-invariant phase difference, $k$ and $l$ signify coordinate directions, and $\mathbf{k}$ is a vector pointing from a lattice site to the next site in the $k$-direction. We use periodic boundary conditions in all three spatial directions. The thermal probability distribution for configurations of the system at inverse temperature $\beta$ is given by the Boltzmann weight
\begin{equation}
  \mathrm{e}^{-\beta F}, \quad F = h^3 \sum_{\mathbf{r}} f(\mathbf{r}),
\end{equation}
and we generate representative samples from these thermal distributions using Monte Carlo simulation.

We now describe the quantities that are measured during the simulations and the methods we use to locate phase transitions.

\subsection{Locating superconducting transitions}
Superconducting transitions in zero external magnetic field can be located using the dual stiffness \cite{Motrunich2008, Herland2013, Carlstrom2015}
\begin{equation}
  \rho^\mu(\mathbf{q}) =
    \left\langle \frac{\left| \sum_{\mathbf{r},\nu,\lambda}
      \epsilon_{\mu\nu\lambda} \Delta_\nu A_\lambda(\mathbf{r})
      \mathrm{e}^{\mathrm{i}\mathbf{q} \cdot \mathbf{r}} \right|^2}
      {(2\pi)^2 L^3} \right\rangle,
\end{equation}
where $\epsilon_{\mu\nu\lambda}$ is the Levi-Civita symbol, $\Delta_\nu$ is a difference operator and $\langle \cdot \rangle$ is the thermal expectation value. In particular, we consider the dual stiffness in the $z$ direction evaluated at the smallest relevant wave vector in the $x$ direction $\mathbf{q}_\mathrm{min}^x = (2\pi/L, 0, 0)$, i.e.\ $\rho^z(\mathbf{q}_\mathrm{min}^x)$, which we denote simply as $\rho$. In the thermodynamic limit, this quantity is zero in the superconducting phase in which fluctuations of the magnetic field are suppressed, and non-zero in the normal phase. Thus, it is a dual order parameter in the sense that it is zero in the low-temperature phase and non-zero in the high-temperature phase. At the critical point of a continuous superconducting transition, the quantity $\rho$ is expected to scale as $1/L$, so that $L\rho$ is a universal quantity. We use finite-size crossings of $L\rho$, extrapolated to the thermodynamic limit, in order to locate superconducting transitions.

\subsection{Locating $Z_2$ phase transitions}
Time-reversal symmetry is $Z_2$, and is thus described by the previously defined Ising order parameter $m$. In order to locate transitions to states that break time-reversal symmetry, we use the Binder cumulant \cite{Binder1981a, Binder1981b} for the order parameter $m$:
\begin{equation}
  U = \frac{\langle m^4 \rangle}{3\langle m^2 \rangle^2}.
  \label{binder_cumulant}
\end{equation}
In the thermodynamic limit, this quantity is equal to $1$ in the high-temperature phase in which the distribution function for the order parameter is given by a single Gaussian centered around $m = 0$, and equal to $1/3$ in the low-temperature phase in which the distribution function has two separate peaks at $m = \pm m_0 \neq 0$. For a continuous transition, the distribution function for the order parameter is expected to have a universal shape at the critical point, so that the Binder cumulant $U$ is a universal quantity. We use finite-size crossings of the Binder cumulant $U$, extrapolated to the thermodynamic limit, in order to locate $Z_2$ transitions.

\subsection{External magnetic field}
In order to implement external magnetic field, we write the vector potential as a sum of two terms: $\mathbf{A}(\mathbf{r}) = \mathbf{A}_0(\mathbf{r}) + \mathbf{A}_1(\mathbf{r})$. The first term corresponds to a uniform magnetic field in the $z$ direction, implemented in the Landau gauge, and is held fixed: $\mathbf{A}_0(\mathbf{r}) = (0, 2\pi x f, 0)$. The second term $\mathbf{A}_1(\mathbf{r})$ is allowed to fluctuate thermally. Due to the periodic boundary conditions imposed on $\mathbf{A}_1(\mathbf{r})$, the contribution to the total magnetic flux from this part of the vector potential is zero, so that the total flux is constant and equal to that given by $\mathbf{A}_0(\mathbf{r})$. In order to be consistent with the periodic boundary conditions, the value of $f$ must be chosen so that $eLh^2f$ is an integer.

In order to detect the presence or absence of a vortex lattice in external magnetic field, we consider two types of quantity: vorticities (to be defined below) and magnetic flux density. For both types of quantity, we measure averages over the direction of the applied field (the $z$ direction), thus obtaining 2D images. We also consider the absolute values of the Fourier transforms of these, which are structure factors.

The vorticity is defined as follows: For a phase field defined on continuous space, the meaning of there being a vortex at a certain point is clear: the phase winding around this point is nonzero. For a phase field defined only on a discrete lattice the meaning of there being a vortex is less clear, and a definition that is reasonable and consistent with the continuum limit must be made. The standard way to count the number of vortices on a given plaquette is this: For each link of the plaquette, consider the gauge-invariant phase difference $\chi_{i,k}(\mathbf{r})$. For each multiple of $2\pi$ that must be added (subtracted) to $\chi_{i,k}(\mathbf{r})$ in order to bring it into the primary interval $(-\pi, \pi]$, add $+1$ ($-1$) to the vorticity of the plaquette.

\subsection{Assessing the existence of the quartic metal phase in the minimal Ginzburg-Landau model in the type~II regime in zero external field}
\begin{figure}
  \centerline{
    \includegraphics{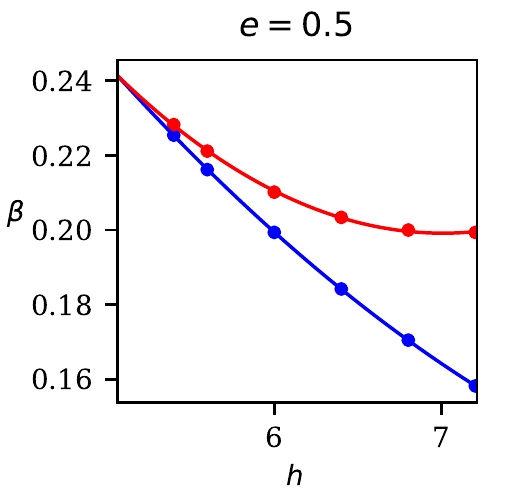}
	  \includegraphics{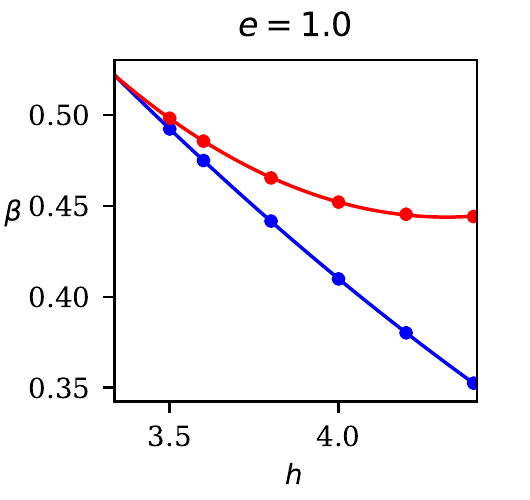}
	  \includegraphics{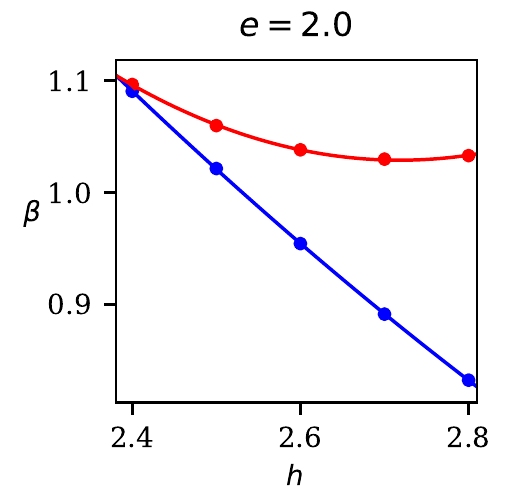} }
	\vspace{8pt}
  \centerline{
    \includegraphics{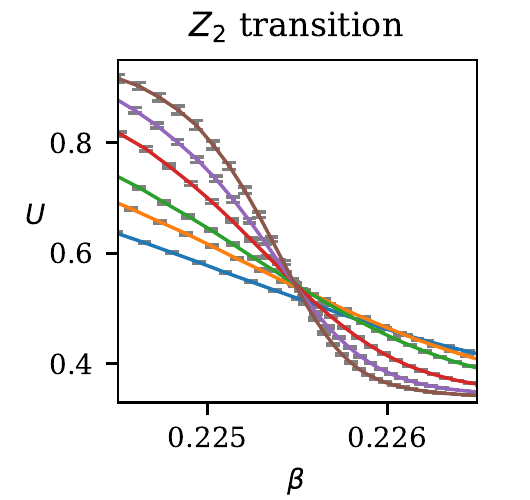}
	  \includegraphics{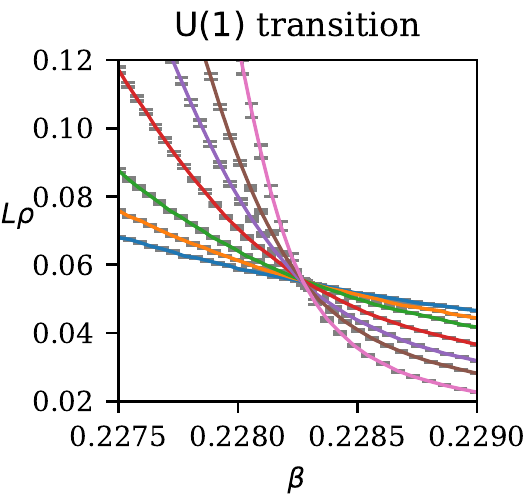}
    \includegraphics{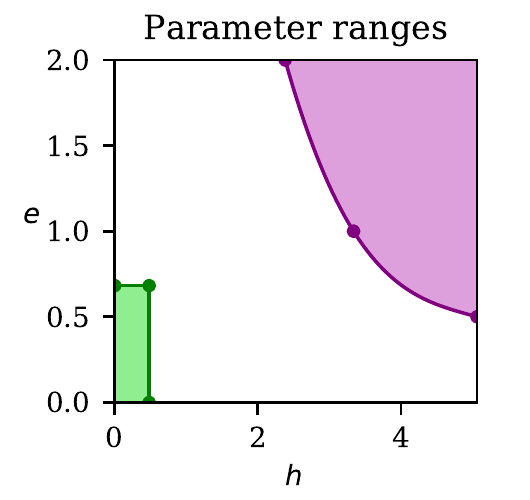} }
  \caption{\textbf{Top row:} Phase diagrams in terms of inverse temperature $\beta$ and lattice constant $h$ for the three charges $e = 0.5, 1.0$, and 2.0 with $Z_2$ transitions shown in blue (bottom curves) and superconducting transitions in red (top curves). The lines are second-order polynomials fitted to the transition points; we use these lines to estimate the positions of the points where the transitions merge. Errors are estimated to be smaller than symbol sizes. \textbf{Bottom row, left and center:} Examples of how the points in the above phase diagrams are determined using finite-size crossings of the Binder cumulant $U$ and the quantity $L\rho$. The parameters are $e = 0.5$, $h = 5.4$ (leftmost pair of points in the above diagrams). The system sizes are $L = 8$ (blue), $10$, $12$, $16$, $20$, and $24$ (brown) for the $Z_2$ transition; for the $\mathrm{U}(1)$ transition $L = 32$ (pink) is also shown. \textbf{Bottom row, right:} Regions of $e$-$h$ space for which the quartic bosonic $Z_2$-metal phase is estimated to occur (purple, top right; the line is a guide to the eye), and for which the system is type II and the lattice constant is the shortest length scale (green, bottom left). }
  \label{phase_diags}
\end{figure}

We have performed Monte-Carlo simulations of the model we consider for various values of the electric charge $e$ and the lattice constant $h$. We consider three values of the electric charge: $e = 0.5, 1.0$ and 2.0. We note that these values correspond to the cases where the magnetic-field penetration length is the largest, an intermediate and the smallest length scale, respectively. For each value of the charge, we construct a phase diagram in terms of inverse temperature $\beta$ and lattice constant $h$, as shown in Fig.~\ref{phase_diags}. The phase transitions are located by considering finite-size crossings of the Binder cumulant $U$, and of the dual stiffness $\rho$ scaled by system size $L$, examples of which are also shown in Fig.~\ref{phase_diags}.

Using the aforementioned phase diagrams, we estimate for which values of the charge $e$ (which in these units is a parameter that sets the magnetic-field penetration length) and the lattice constant $h$ the anomalous fluctuation-induced quartic metal phase occurs. We do this by fitting second-order polynomials to the determined points on the phase diagrams in Fig.~\ref{phase_diags}, in order to estimate the positions of the bicritical points at which the transitions merge. The result is shown in Fig.~\ref{phase_diags}. For the system sizes and lattice discretization that we consider, the quartic metal phase is estimated to occur for values of $e$ and $h$ greater than those indicated by the purple dots, thus roughly in the purple region in the top-right corner (the line is a guide to the eye).

In order to assess if the material is described by the minimal version of the model 
we started from, with some fluctuation corrections, we consider the following two requirements. First, the indications are that the material is relatively strongly type-II when substantially outside of the $Z_2$-phase transition. Thus, in setting up the lattice model, the bare penetration length must be large enough relative to the other characteristic length scales. In our dimensionless units, the electric charge parameterizes the magnetic-field penetration length, and thus the electric charge must be small enough. Second, the lattice constant must be small enough relative to the other characteristic length scales. The region of $e$-$h$ space for which the model is type II and the lattice constant is the shortest length scale is that shown in green in Fig.~\ref{phase_diags}. The green (relevant parameters) and purple (quartic metal) regions are well separated, which indicates that the quartic metal phase does not
occur in the type-II regime of our lattice realiazation   of this minimal continuum Ginzburg-Landau model in the absence of applied field, for the lattice sizes that we considered. Since the experimental results indicate that the quartic metal phase does exist in zero field, this suggests that this system may 
be better described by lattice Londin model with fluctuations \cite{Bojesen2013,Bojesen2014a}.
This is fully consistent with the Nernst effect for samples with $x = 0.8$ that shows pairing fluctuations at temperatures approximately twice as high as the superconducting $T_c$. The conclusion we can draw is 
that higher-order terms which enhance the phase-difference stiffness relative to the phase-sum stiffness are indeed important. These terms change the domain-wall energy compared to the vortex energy, and thus should create the quartic metal phase in zero field. 
\subsection{Assessing the existence of the quartic metal phase in the  presence of mixed gradient terms}
We next turn to a more general model that includes mixed-gradient terms. Such terms are allowed by symmetry and are generically present in multicomponent systems.
Since the coefficients of these terms have not been derived microscopically for iron-based superconductors, we add them phenomenologically. As discussed in Refs.~\cite{Dahl2008a, Herland2010}, mixed-gradient terms cannot be discretized using the discretization schemes used  for simpler Ginzburg-Landau models. To treat these terms we have to resort to the Villain representation of the London model as described below. In this section, we consider the reduced two-component London model~\eqref{2c_london} in the limit of infinite penetration length. The two-component model is obtained by projecting a three-band model onto two fields that correspond to linear combinations of the gap fields in the three bands \cite{Garaud2017}. As mentioned above, these approximations underestimate fluctuations and the presence of $Z_2$ phase. Starting from the functional Eq.~\eqref{2c_london}, we consider the case $\rho_1 = \rho_2 = \rho$. By rescaling the coupling constants and the free energy according to:
\begin{equation*}
    \nu \mapsto \nu \rho , \quad \eta_2 \mapsto \eta_2 \rho, \quad f \mapsto f/ \rho,
\end{equation*}
we can reduce the number of free parameters in the model, so that the resulting free-energy density reads:
\begin{equation}
    f = \sum_{i=1}^2 \frac{1}{2} \left( \mathbf{\nabla} \phi_i \right)^2 -\nu \left(\mathbf{\nabla} \phi_1 \cdot \mathbf{\nabla}\phi_2 \right) + \eta_2 \cos[2(\phi_1 -\phi_2 )].
    \label{final_continuum_London}
\end{equation}
To highlight the role of the mixed-gradient term, we can conveniently rewrite Eq.~\eqref{final_continuum_London} in terms of the inter-component phase-sum and phase-difference modes as:
\begin{equation}
f= \frac{1 - \nu}{4} \left( \mathbf{\nabla} \phi_1 + \mathbf{\nabla} \phi_2\right)^2 + \frac{1 +\nu}{4} \left(\mathbf{\nabla} \phi_1 - \mathbf{\nabla} \phi_2 \right)^2  + \eta_2 \cos[2(\phi_1 -\phi_2 )].
\label{final_continuum_London2}
\end{equation}
Increasing the value of $\nu$, the bare stiffness of the phase-sum mode, associated with the superconducting $U(1)$ symmetry, decreases. Conversely, the bare stiffness of the phase-difference mode, associated with the $Z_2$ symmetry, increases. From Eq.~\eqref{final_continuum_London2}, it is straightforward to derive the stability condition for the model (i.e.\ that the free-energy functional is bounded from below), which is simply $\nu < 1$.

\subsection{Villain representation of London model of s+is superconductor with mixed gradient terms}
To perform Monte Carlo simulations, we need to provide a discrete lattice representation of the continuum model Eq.\eqref{final_continuum_London}. A faithful discretization scheme that allows an artifact-free representation of the mixed gradient terms is the Villain approximation~\cite{Villain1975}, which accommodates the compactness of the phase by rewriting:
\begin{equation*}
    e^{\beta \cos{ \left( \Delta_{\mu} \phi_i \right)} }  \to \sum_{n=-\infty}^{\infty} e^{-\frac{\beta}{2} (\phi_{i+\mu} - \phi_i- 2\pi n)^2},
\end{equation*}
where we have fixed the value of the lattice spacing to $h = 1$. The Villain Hamiltonian for the model~\eqref{final_continuum_London} reads:
\begin{equation}
        H_v= \sum_{r, \mu} V_{\mu}(\Delta_{\mu} \phi_1,\Delta_{\mu} \phi_2, \phi_1, \phi_2; \beta)= -\sum_{r, \mu} \beta^{-1} \ln \left\{ \sum_{n_{1,\mu} n_{2,\mu}} e^{-{\beta} S} \right\} ,
        \label{H_Vill}
\end{equation}
where
\begin{equation}
\begin{split}
    S = \frac{1}{2} [(\Delta_{\mu} \phi_1 -2\pi n_{1, \mu})^2 &+(\Delta_{\mu} \phi_2 -2\pi n_{2, \mu})^2] - \nu\big[\Delta_{\mu}( \phi_1 -\phi_2)+ \\&-2\pi(n_{1, \mu} -n_{2, \mu})\big]^2 + \eta_2 \cos[2(\phi_1 -\phi_2 )].
\end{split}
\end{equation}
Finally, we set the value of the Josephson coupling constant to the relatively small value $\eta_2 = 0.1$, in order to ensure that the lattice spacing $h$ is the smallest length scale at play. Increasing the Josephson coupling increases the size of the $Z_2$ phase. By expanding the free energy around the ground state, one indeed finds that the characteristic length scale at which the perturbed phase-difference recovers its ground state value is $\lambda_J= \sqrt{\frac{1 + \nu}{ 8 \eta_2}}$. The choice $\eta_2 = 0.1$ guarantees $h < \lambda_J$ for all possible values of $\nu$.

We have performed Monte Carlo simulations of the Villain Hamiltonian Eq.~\eqref{H_Vill}, locally updating the two fields $\phi_1, \phi_2 \in [0, 2\pi )$ by means of the Metropolis-Hastings algorithm. We have considered a three-dimensional system with $L^3$ sites for different values of the linear size $L$, as is needed to properly assess the critical points of the model.

\subsection{Locating the $U(1)$ and $Z_2$ transitions within the two-component London model}
The $U(1)$ transition is associated with the onset of the superfluid phase, which is captured by the helicity modulus of the phase sum \cite{Dahl2008a, Herland2010}. In a multicomponent system, one can define several helicity moduli corresponding to different linear combinations of individual phases. In the two-component case, for each choice of the coefficients $\{ a_i \}$ in
\begin{equation}
\begin{pmatrix} \phi'_1(\mathbf{r}) \\  \phi'_2(\mathbf{r})  \end{pmatrix}
= \begin{pmatrix} \phi_1(\mathbf{r}) \\  \phi_2(\mathbf{r})  \end{pmatrix} 
+  \begin{pmatrix} a_1 \\  a_2  \end{pmatrix} \mathbf{\delta} \cdot \mathbf{r},
\end{equation}
one can define a corresponding helicity modulus
\begin{equation}
\Upsilon_{\mu, \{a_i\}} =\frac{1}{N} \frac{\partial^2 F(\{\phi'_i\}) }{\partial \delta_{\mu}^2}\Bigr|_{\delta_{\mu}=0} = \sum_{i} a_i^2 \Upsilon_{\mu, i} + 2\sum_{i<j} a_i a_j \Upsilon_{\mu, ij},
\label{Helicity_multicomponent1}
\end{equation}
where
\begin{eqnarray}
\label{Helicity1}
 \Upsilon_{\mu, i} &=&   \frac{1}{N} \left[ \Big\langle \frac{\partial^2 H}{ \partial \delta_{\mu,i}^2}
  \Big\rangle - \beta \Big\langle \left(  \frac{\partial H}{ \partial \delta_{\mu,i}}  - \langle \frac{\partial H}{ \partial \delta_{\mu,i}} \rangle \right)^2  \Big\rangle  \right]_{\delta_{\mu,i }=0} \\
 \label{Helicity2}
 \Upsilon_{\mu, ij}&=&  \frac{1}{N} \left[ \Big\langle \frac{\partial^2 H}{ \partial \delta_{\mu,i} \partial \delta_{\mu,j} } \Big\rangle - \beta \Big\langle \left(  \frac{\partial H}{ \partial \delta_{\mu,i}} - \langle \frac{\partial H}{ \partial \delta_{\mu,j}} \rangle \right)^2  \Big\rangle \right]_{\delta_{\mu, ij}=0}.
\end{eqnarray}
For our purposes, the relevant observable is the phase-sum helicity modulus $\Upsilon_{+}$ defined by the choice $a_1 = a_2 = 1$. We determine the critical temperature associated with the $U(1)$ transition by using finite-size crossings of the quantity $L \Upsilon_{+}$. In Fig.~\ref{FigED3}-(c) we report the finite-size crossings of $L \Upsilon_{+}$ for $\nu = 0.6$ and $L = 8, 10, 12, 16, 20, 24, 32$.

To locate the $Z_2$ transition, we define an Ising order parameter $m$ (similar to that used in the three-component case) distinguishing between the two possible chiralities [see Fig.~\ref{FigED3}-(d)] and  consider the finite-size crossings of the associated Binder cumulant $U$ defined in Eq.~\eqref{binder_cumulant}. In Fig.~\ref{FigED3}-(b) we report the finite-size crossings of $U$ for $\nu = 0.6$ and $L = 8, 10, 12, 16, 20, 24, 32$.

The aforementioned results show the presence of a quartic metal phase in three dimensions in zero external magnetic field in extreme type-II limit, when one goes beyond the simplest three-component Ginzburg-Landau model. A separate and more extended theoretical study of the model given by Eq.~\eqref{2c_london} will be published elsewhere \cite{Illaria}.

\subsection{Quartic metal phase in non-zero field}
We show that the quartic metal phase occurs in the minimal Ginzburg-Landau model we consider in non-zero field. At the level of London models, it has previously been demonstrated that this kind of phase forms in external field because melting of the lattice of one-quanta vortices restores the $\mathrm{U}(1)$ symmetry but preserves the broken symmetry associated with the phase differences of the order parameters \cite{babaev2004superconductor, Smorgrav2005b, Bojesen2014b}. We demonstrate that this effect is present in the minimal Ginzburg-Landau model (\ref{contf2}). Specifically, we give an example of this for the charge $e = 0.5$, which makes the system type II, and the lattice constant $h = 0.7$, which is shorter than the penetration depth, the characteristic length scale of the lightest phase-density mode (that with the longest length scale) and the length scale of the total density mode. (In fact, the two lightest phase-density modes are degenerate, as are the two heaviest ones.) The system size used is $L = 64$, and the applied magnetic field is 
$H = 2\pi/(eLh^2)$.

\bibliography{arxiv2}

\end{document}